\documentclass[fleqn,usenatbib]{mnras}

\usepackage{newtxtext,newtxmath}

\usepackage[T1]{fontenc}

\DeclareRobustCommand{\VAN}[3]{#2}
\let\VANthebibliography\thebibliography
\def\thebibliography{\DeclareRobustCommand{\VAN}[3]{##3}\VANthebibliography}

\usepackage{graphicx}	
\usepackage{color}
\usepackage{amsmath}	

\usepackage{ulem}

\definecolor{mmr}{rgb}{0.7,0.0,0.2}

\definecolor{mmb}{rgb}{0.0,0.3,1.0}

\definecolor{mmg}{rgb}{0.0,0.7,0.3}

\title[Chemical abundances of symbiotic giants]{Chemical abundance analysis of symbiotic giants.\\ Metallicity and CNO abundance patterns in 14 northern S-type systems}

\author[C. Ga{\l}an et al.]{
Cezary Ga{\l}an,$^{1}$\thanks{E-mail: cgalan@camk.edu.pl (CG)}
Joanna Miko{\l}ajewska,$^{1}$
Kenneth H. Hinkle,$^{2}$
Richard R. Joyce$^{2}$
\\
$^{1}$Nicolaus Copernicus Astronomical Center, Polish Academy of Sciences, Bartycka 18, PL-00-716 Warsaw, Poland\\
$^{2}$NSF's National Optical-Infrared Astronomy Research Laboratory, 950 N. Cherry Ave., Tucson, AZ 85719, USA
}

\date{Accepted 2023 July 8. Received 2023 July 7; in original form 2022 May 9}

\pubyear{2023}

\begin{document}
\label{firstpage}
\pagerange{\pageref{firstpage}--\pageref{lastpage}}
\maketitle

\begin{abstract}
In previous works, we computed abundances for the red giant in nearly four
dozen S-type symbiotic systems (SySt).  The abundances provide information
about metallicity, evolutionary status, and possible memberships in Galactic
stellar populations.  
Here, we extend our studies with a northern hemisphere sample of SySt.  This
northern sample is dominated by Galactic disc/halo objects, whereas our
previous southern sample is heavily biased toward the bulge population.  
Spectrum synthesis of high-resolution (R$\sim$50\,000), near-$IR$ spectra
using standard LTE analysis and atmospheric models have been used to measure
abundances of CNO and elements around the iron peak (Fe, Ti, Ni, and Sc) in the
atmospheres of the red giant component. 
The SySt sample shows generally slightly sub-solar metallicity, as expected
for an older disc population, with a median at [Fe/H]\,$\sim -0.2$\,dex. 
Enhanced $^{14}$N, depleted $^{12}$C, and decreased $^{12}$C/$^{13}$C
indicate that all these giants have experienced the first dredge-up. 
Comparison with theoretical predictions indicates that additional mixing
processes had to occur to explain the observed C and N abundances.  Relative
O and Fe abundances agree with those represented by Galactic disc and bulge
giant populations in the {\sl APOGEE} data, with a few cases that can be
attributed to membership in the extended thick-disc/halo.  As an interesting
byproduct of this study, we observed a blue-shifted additional component on
the wings of absorption lines in the spectra of AG\,Peg which could be
connected with accretion on to the hot component.
\end{abstract}

\begin{keywords}
stars: abundances -- stars: atmospheres -- binaries: symbiotic -- stars: evolution -- stars: late-type.
\end{keywords}



\section{Introduction}

Stellar astrophysics is now able to provide a fairly complete picture of
single star evolution.  However, a number of unresolved issues remain
concerning binary evolution, particularly with symbiotic stars (SySt). 
These systems are composed of strongly interacting stars at remarkably
different stages of evolution: an evolved red giant (RG) donor (either a
normal giant in S-type SySt, or a Mira variable embedded in an optically
thick dust envelope in D-type SySt) which transfers material to a hot,
luminous companion, typically a white dwarf (WD) but a main-sequence star
with an accretion disc or neutron star is also suggested in some cases,
surrounded by an ionized nebula.  SySt represent the interacting binary
systems with the longest orbital periods ($P_{\rm{orb}} \sim$ years to
centuries) with separations large enough to accommodate the RG branch
(RGB) or asymptotic giant branch (AGB) stars.  SySt offer insight into all
interacting binaries that include evolved RGB/AGB stars and accreting WDs
during any phase of their evolution \citep[see][for a recent reviews of
SySt]{Mik2012, Mun2019}.

Since some of SySt contain a massive WD with a high accretion rate, they
have been proposed as possible progenitors for type Ia supernovae \citep[see
e.g.][]{Hac1999, MiSh2017, Liu2018}.  A review of the most promising SN Ia
progenitors among SySt is presented in \citet{Mik2013}.

The mass exchange between the components of binary systems is critical in
defining their evolution.  The symbiotic giant can range from spectral type
G to late M, with corresponding differences in binary separation and mass
transfer.  The symbiotic giants are losing matter at a rate ($\geq
10^{-7}$Myr$^{-1}$) which is systematically higher than for single field
giants \citep[see][]{Mik2003b}.  A substantial part of this mass can be
accreted on to the compact object from wind and/or via Roche lobe overflow
\citep{Mik2012}.  When the system was formed, the current hot stellar
remnant was the more massive component, which when passing through its RG
stage transferred part of its mass to the main-sequence companion, that is
currently RG.  That mass transfer episode should have left traces in the RG
chemical composition, and indeed such chemical pollution has been detected
in some RG-WD binary systems \citep{SmLa1988}.  In SySt with circularized
orbits characterized by periods shorter than
$P_{\rm{orb}}$\,$\simeq$\,900\,days one can expect that in most of them the
mass transfer must have taken place in the past \citep[see Section 2
of][]{Mik2012}.\\
\indent To study the issues of binary interactions and evolution in SySt, as
well as their population origin, the chemical composition of the symbiotic
giant atmospheres provides additional insight.  Useful diagnostics include
abundances of some specific chemical elements; e.g.  s-process elements
produced during the AGB phase, CNO abundances which can provide information
about evolutionary status, or $\alpha$ elements in relation to metallicity,
which can inform about Galactic population membership.  Our previous
measurements and analyses concerned 37 objects from the southern hemisphere
\citep{Gal2016, Gal2017}, and have been largely focused on the bulge
population.  In this paper, we present abundances of CNO and elements around
the iron peak (Fe, Ti, Ni, and Sc) derived through the spectral synthesis of
the high-resolution near-$IR$ spectra for an additional sample of 14 RGs in
S-type SySt located in the northern hemisphere.  The northern sample is
important as it is dominated by Galactic disc/halo objects, whereas the
southern sample is heavily biased toward the bulge.

The order of the paper is as follows.  The spectroscopic observations and
reductions are presented in Section\,\ref{sc-OaR}.  In Section\,\ref{sc.pg}
we estimate the atmospheric parameters of symbiotic giants.  The applied
methods and finally calculated abundances are shown in
Section\,\ref{sc.AaR}.  The discussion of the results and conclusions are
presented in Section\,\ref{sc.dis} with a comparison to abundances from the
{\sl APOGEE} mass survey and theoretical models.


\section{Observations and data reduction} \label{sc-OaR}

High-resolution (R = $\lambda$/$\Delta \lambda \sim$ 50\,000, S/N $\sim$
100), near-$IR$ spectra of the program stars were observed with the Phoenix
cryogenic echelle spectrograph \citep[see][for a complete
description]{Hin1998} using the 4\,m telescope at the Kitt Peak National
Observatory.  14 SySt were observed during four nights spanning
September 3 -- 6, 2014.  The journal of our spectroscopic observations is
given in Table\,\ref{T-js}.  Three narrow spectral intervals were covered. 
One $H$-band region (width of $\sim$\,65\,\AA) at a mean wavelength close to
15635\,\AA, and two slightly wider regions ($\sim$ 100 \AA) around
22275\,\AA, and 23635\,\AA, hereafter designated as $K$- and $K_{\rm
r}$-band spectra, respectively.  In most cases, all three regions were
explored; however, for the five objects T\,CrB, FG\,Ser, V443\,Her,
V1413\,Aql, and CH\,Cyg only the $H$-band region was observed.

The spectra were extracted from the raw data and wavelength
calibrated using standard reduction techniques \citep{Joy1992}.  The
left and right aperture data for each object were reduced separately and
combined into a single spectrum.  In line with common practice, the
wavelength scales of all spectra were heliocentric corrected.  Telluric
lines were removed (except for the $H$-band region that is free of telluric
features) by reference to a hot standard star, that was observed at
approximately the same time.  The Gaussian instrumental profile in all cases
is $\sim$\,6\,km\,s$^{-1}$ full width at half-maximum (FWHM), corresponding
to instrumental profiles of $\sim$ 0.31, 0.44, and 0.47\,\AA\, in the case
of the $H$-, $K$- and $K_{\rm r}$ -band spectra, respectively.  

\begin{figure}
   \centering
   \includegraphics[width=\hsize]{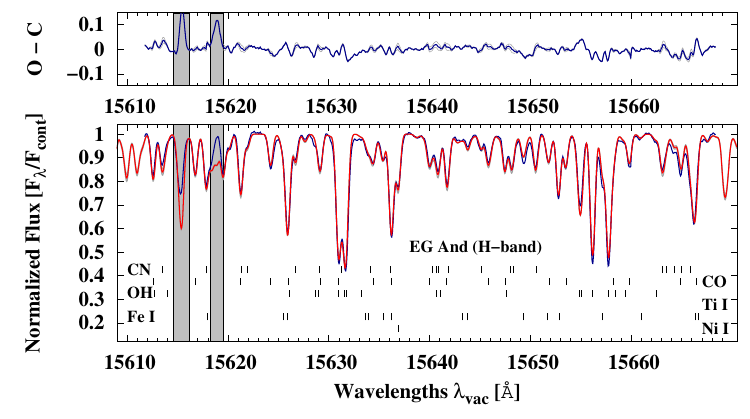}
      \caption{The $H$ band spectrum of EG\,And (blue line) and a synthetic
              spectrum (red line) calculated using the final abundances
              (Table\,\ref{T-finAbu}). The grey-shaded areas were
              excluded from calculations by a suitable mask.}
         \label{F1}
   \end{figure}

\begin{figure}
   \centering
   \includegraphics[width=\hsize]{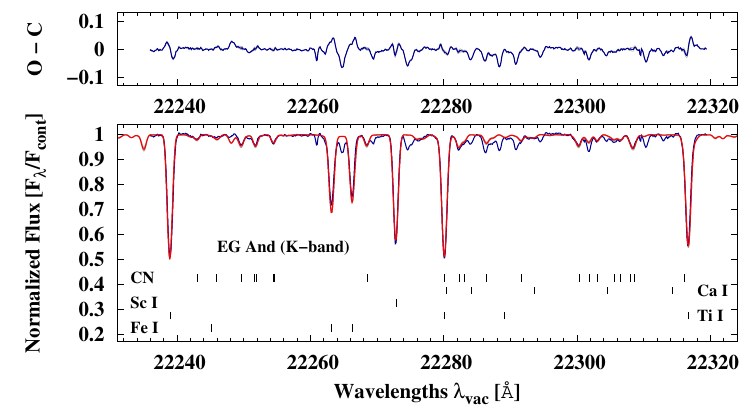}
      \caption{The $K$ band spectrum of EG\,And (blue line) and a synthetic
              spectrum (red line) calculated using the final abundances (Table\,\ref{T-finAbu}).}
         \label{F2}
   \end{figure}

\begin{figure}
   \centering
   \includegraphics[width=\hsize]{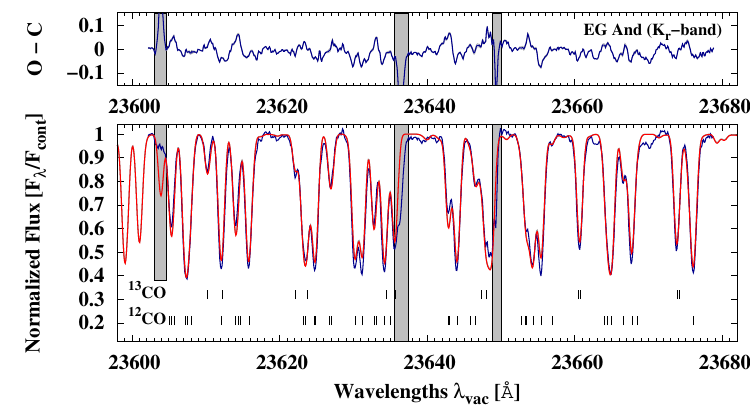}
      \caption{The $K_{\rm r}$ band spectrum of EG\,And (blue line) and a
              synthetic spectrum (red line) calculated using the final
              abundances (Table\,\ref{T-finAbu}). The grey-shaded areas
              were excluded from calculations by a suitable mask.}
         \label{F3}
   \end{figure}

Example spectra of EG\,And in all three intervals are shown in
Figs\,\ref{F1} -- \ref{F3}.  $H$-band spectra contain relatively strong
first overtone OH lines and a selection of neutral atomic lines from
\ion{Fe}{i}, \ion{Ti}{i}, \ion{Ni}{i} that are superimposed on background
weaker second-overtone CO vibration-rotation lines and CN red system $\Delta
\nu$ = $-$1\,CN lines.  The $K$-band region contains moderately strong Ti I
lines as well as a few other neutral atomic lines from Fe I and Sc I
superimposed on weak CN molecular lines from the CN red system $\Delta \nu$
= $-2$ transition.  The molecular lines from these two regions were used to
determine abundances of C, N, and O.  The atomic lines were used to derive
abundances of elements around the iron peak: Sc, Ti, Fe, Ni.  The $K$r-band
interval is dominated by strong CO features that are heavily blended.  These
were used to measure the $^{12}$C/$^{13}$C isotopic ratio.

\begin{table}
 \centering 
  \caption{Journal of spectroscopic observations obtained at $H$-
($\sim$15600\,\AA), $K$- ($\sim$22300\,\AA), and $K_{\rm r}$-band
($\sim$23600\,\AA) regions with Phoenix spectrograph during 4 nights
(September 3--6, 2014).  The order of targets is according to the increasing
R.A.  -- here as well as later in the tables and in appendixes.  Orbital
phases have been calculated according to the referenced literature
ephemeris.  In the right columns are the values of rotational velocity
measured from our spectra (see Section\,\ref{sc-OaR}).  The velocity unit is
${\rm{km}}$\,${\rm{s}^{-1}}$.  Rotational velocities have been obtained from
the measurement of full width at half-maximum (FWHM) of band \ion{Ti}{i},
\ion{Fe}{i}, and \ion{Sc}{i} absorption lines.  The last column presents the
value of rotational velocity obtained from the fit of synthetic spectra.}
\label{T-js}
  \begin{tabular}{@{}l@{\hskip 5mm}c@{\hskip 5mm}c@{\hskip 5mm}c@{\hskip 5mm}c@{\hskip 5mm}c@{}}
             & Sp.\,reg.   &  HJD(mid) & Phase$^{a}$ & \multicolumn{2}{c}{$V_{rot} \sin{i}$} \\
             & band        & -2456900  &             & FWHM         & fit                    \\
 \hline
 EG\,And     & $H$         &  3.95295 & 0.843       &  --          & ~4.9                    \\
             & $K$         &  4.95454 & 0.845       &  5.5$\pm$1.1 & ~5.85                   \\
             & $K_{\rm r}$ &  5.94766 & 0.847       &  --          & ~5.4                    \\
 AX\,Per     & $H$         &  3.95620 & 0.709       &  --          & ~8.2                    \\
             & $K$         &  4.95957 & 0.710       &  9.1$\pm$1.0 & ~9.3                    \\
             & $K_{\rm r}$ &  5.95981 & 0.712       &  --          & ~8.8                    \\
 T\,CrB      & $H$         &  6.66832 & --          &  --          & ~7.1                    \\
 FG\,Ser     & $H$         &  3.74483 & 0.270       &  --          & ~9.3                    \\
 V443\,Her   & $H$         &  3.75275 & 0.189       &  --          & ~5.1                    \\
 V1413\,Aql  & $H$         &  3.76963 & 0.618       &  --          & 16.8                    \\
 BF\,Cyg     & $H$         &  3.80657 & 0.275       &  --          & ~8.0                    \\
             & $K$         &  6.82913 & 0.279       &  8.5$\pm$0.9 & ~8.9                    \\
             & $K_{\rm r}$ &  5.83552 & 0.278       &  --          & ~8.8                    \\
 CH\,Cyg     & $H$         &  3.84599 & 0.855       &  --          & ~8.1                    \\
 QW\,Sge     & $H$         &  3.81973 & --          &  --          & ~7.2                    \\
             & $K$         &  6.84208 & --          &  8.4$\pm$1.1 & ~8.6                    \\
             & $K_{\rm r}$ &  5.84895 & --          &  --          & ~8.1                    \\
 CI\,Cyg     & $H$         &  3.83156 & --          &  --          & 11.5                    \\
             & $K$         &  6.85385 & --          & 12.3$\pm$0.4 & 12.8                    \\
             & $K_{\rm r}$ &  5.86198 & --          &  --          & 12.2                    \\
 PU\,Vul     & $H$         &  3.83743 & --          &  --          & ~6.5                    \\
             & $K$         &  6.85975 & --          &  7.5$\pm$1.0 & ~8.4                    \\
             & $K_{\rm r}$ &  5.86988 & --          &  --          & ~8.4                    \\
 V1329\,Cyg  & $H$         &  3.91592 & 0.582       &  --          & ~9.5                    \\
             & $K$         &  4.92762 & 0.583       & 10.3$\pm$0.5 & 10.1                    \\
             & $K_{\rm r}$ &  5.92546 & 0.584       &  --          & ~9.8                    \\
 AG\,Peg     & $H$         &  3.93592 & --          &  --          & ~7.9                    \\
             & $K$         &  4.94291 & --          &  8.2$\pm$1.1 & 10.0                    \\
             & $K_{\rm r}$ &  5.93767 & --          &  --          & ~9.7                    \\
 Z\,And      & $H$         &  3.93949 & 0.753       &  --          & ~5.9                    \\
             & $K$         &  4.94679 & 0.755       &  7.5$\pm$0.9 & ~7.9                    \\
             & $K_{\rm r}$ &  5.94139 & 0.756       &  --          & ~6.2                    \\
 \hline
  \end{tabular}   
\begin{list}{}{}
\item[{\sl {Notes.}} $^{a}$Orbital]\,phases are calculated from
the following ephemerides: EG\,And $2450683.16+482.57\times$E
\citep{KeGa2016}; AX\,Per $2450963.8 + 682.1\times$E \citep{Fek2000b};
FG\,Ser $2451031.4 + 633.5\times$E \citep{Fek2000b}; V443\,Her $2450197.3 +
599.4\times$E \citep{Fek2000b}; V1413\,Aql $2450567.11 + 433.47\times$E
\citep{Poy2012}; BF\,Cyg $2451395.2 + 757.2\times$E \citep{Fek2001}; CH\,Cyg
$2446353 + 5689\times$E \citep{Hin2009}; V1329\,Cyg $2451565.0 +
956.5\times$E \citep{Fek2001}; Z\,And $2450260.2 + 759.0\times$E
\citep{Fek2000b}.]
\end{list}
\end{table}

\section{Red giant parameters}  \label{sc.pg}

Calculation of chemical composition requires the use of model atmospheres
corresponding to specific parameters characterizing the physical conditions
prevailing in stellar atmospheres.  The most important are effective
temperature ($T_{\rm{eff}}$) and surface gravity ($\log{g}$).  They are
predominantly derived based on analysis of spectral features from neutral
and ionized species, usually of Fe.  Lines of ionized species are not common
in the near-infrared ($IR$) and our spectra, covering narrow wavelength
ranges, do not contain any lines from ionized elements.  In addition, there
are not enough unblended lines with sufficiently different intensities for
the same elements to be useful for determining these parameters.  Therefore,
we have deduced the stellar parameters using different methods and
information available in the literature.

%

%

%

Estimation of $T_{\rm{eff}}$ was largely based on spectral types
(Table\,\ref{T-est}) derived by \citet{Mue1999} from analysis of TiO bands
in the near-$IR$ ($I$-band).  However, in the case of AX\,Per,
V1413\,Aql, PU\,Vul, V1329\,Cyg, and AG\,Peg we adopted slightly later
spectral types based on the spectra that were derived very close to inferior
conjunction (RG component in front) or during mid-eclipses \citep[see][and
references therein]{Bel2000} and thus were less influenced by the hot
companion.  Calibrations by \citet{Ric1999} and \citet{VBe1999} were used to
translate the spectral types into $T_{\rm{eff}}$.  Independent estimation of
an upper limit for temperature was derived from $J - K$ colours of Two
Micron All Sky Survey (2MASS) photometry with the use of \citet{Kuc2005}
$T_{\rm{eff}}$--$\log{g}$--colour relation for late-type giants gives
generally consistent results (Table\,\ref{T-est}).  The final values of
$T_{\rm{eff}}$ adopted for further use are listed in the penultimate column
of Table\,\ref{T-est}.

The majority of systems studied here are eclipsing binaries.  In many cases,
parameters of the RG are known (Table\,\ref{T-MaR}) from orbital analysis of
both components (T\,CrB \citep{Sta2004}; AX\,Per \citep{MiKe1992}; BF\,Cyg
\citep{Fek2001}; V1329\,Cyg \citep{Fek2001, Pri2003}; AG\,Peg
\citep{Fek2000a, Ken1993}) or at least of the RG component (EG\,And
\citep{KeGa2016}; FG\,Ser, V443\,Her, Z\,And \citep{Fek2000b}; CH\,Cyg
\citep{Hin2009}) and/or from analysis of ellipsoidal variability in light
curves (EG\,And \citep{WiVa1997}; T\,CrB \citep{BeMi1998}; BF\,Cyg
\citep{Yud2005a}).  In two cases parameters derived from interferometric
observations (FG\,Ser, AG\,Peg \citep{Bof2014}) were also used.  This allows
an estimate of the RG surface gravity (Table\,\ref{T-MaR} and
Table\,\ref{T-elogg} -- column ({\sl{II}})).  In the absence of such
information, the parameters were based solely on the Gaia (DR\,3 and DR\,2)
parallaxes \citep{Gai2023, Gai2021, Gai2018} to estimate the RG's radius
(V1413\,Aql and QW\,Sge).  We relied on the {\sl {gofAL}} (goodness-of-fit
statistic parameter -- see Table\,\ref{T-eBMaRg}) to decide which
measurement adopt as more reliable.  Generally, in most cases, DR\,3 gives a
better fit but in two cases -- V1413\,Aql and BF\,Cyg -- {\sl{gofAL}} is
much smaller in the case of DR\,2, and in these cases, we adopted parallaxes
from this previous release.  However, it must be noted that in the cases of
three targest (see Table\,\ref{T-eBMaRg}) the {\sl {gofAL}} is less than
$\sim$\,3, which according to the Gaia documentation means the good fit to the
data.  Corresponding distances were taken as estimated with the dedicated
special method by \citet{Ba-J2018} which allows obtaining this information
even with really bad quality measurements.  Using known $K_0$ magnitudes and
bolometric corrections ($BC_{\rm{K}}$) by \citet{BeWo1984} we estimated
bolometric magnitudes ($M_{\rm{Bol}}$) and luminosities to derive radii
(Table\,\ref{T-eBMaRg}) and finally calculate $\log{g}$.  The H-R diagram
for objects from our sample compared to the {\sl{BaSTI evolutionary tracks}}
\citep{Hid2018} is shown in Figure\,\ref{HR_A1} in Appendix \ref{sc.npo}.

In cases when we have only spectroscopic orbit information we can use the RG
velocity amplitude, $K_{\rm{g}}$, and its rotational velocity, $V_{\rm{rot}}
\sin{i}$, in combination with Roche lobe geometry to estimate the minimum
mass ratio $q_{\rm{min}}$ using the expression:\\

$(1+q) r(q) \geq \frac{V_{\rm{rot}} \sin{i}}{K_{\rm{g}}}$,\\

\noindent where the Roche limit \citep{Eggl1983,Pac1971}:\\

$r(q) = \frac{0.49 q^{2/3}}{[0.6 q^{2/3} + \ln{(1 + q^{1/3})}]}$\\

\begin{table*}
\caption[]{Estimation of effective temperature $T_{\rm{eff}}$ and $\log{g}$ based on known spectral types and IR colours as well as the values adopted for this study.}
\label{T-est}
{\small{
\begin{tabular}{@{}l@{\hskip 5mm}l@{\hskip 5mm}l@{\hskip 5mm}l@{\hskip 5mm}l@{\hskip 5mm}l@{\hskip 5mm}l@{\hskip 5mm}l@{\hskip 5mm}l@{\hskip 5mm}l@{\hskip 5mm}l@{}}
\hline
            & Sp.T.$^{[1]}$ & $T_{\rm{eff}}^{[2]}$ & $T_{\rm{eff}}^{[3]}$ & $J-K^{[4,5]}$ & $E$($B-V$)$^{[6]}$ & ($J-K$)$_0$      & $T_{\rm{eff}}^{[7]}$ & $\log{g}^{[7]}$ & $T_{\rm{eff}}^{a}$ & $\log{g}^{a}$ \\
            &               & [K]                  & [K]                  & mag           & mag                & mag              & [K]                  &                      & [K]                &               \\
  \hline
 EG\,And    & M3            & 3560\,$\pm$\,75      & 3586                 & $\sim 1.28$   & $<$0.07$\pm$0.01   & $\sim 1.25$      & --                   & --              & 3600               & 0.5           \\
 AX\,Per    & M6$^b$        & 3240\,$\pm$\,75      & 3258                 & 1.29$\pm$0.06 & $<$0.19$\pm$0.01   & $>$1.20$\pm$0.07 & $<$3460$\pm$140      & $<0.3\pm0.2$    & 3300               & 0.0           \\
 T\,CrB     & M4.5          & 3410\,$\pm$\,75      & 3421                 & 1.23$\pm$0.05 & $<$0.06$\pm$0.01   & $>$1.20$\pm$0.06 & $<$3460$\pm$120      & $<0.3\pm0.2$    & 3400               & 0.5           \\
 FG\,Ser    & M5            & 3355\,$\pm$\,75      & 3367                 & 1.56$\pm$0.07 & $<$0.72$\pm$0.03   & $>$1.20$\pm$0.10 & $<$3450$\pm$210      & $<0.3\pm0.4$    & 3400               & 0.5           \\
 V443\,Her  & M5.5          & 3300\,$\pm$\,75      & 3312                 & 1.32$\pm$0.07 & $<$0.13$\pm$0.01   & $>$1.26$\pm$0.07 & $<$3340$\pm$150      & $<0.1\pm0.3$    & 3300               & 0.5           \\
 V1413\,Aql & M4.5$^b$      & 3410\,$\pm$\,75      & 3421                 & 1.27$\pm$0.06 & $<$0.60$\pm$0.01   & $>$0.98$\pm$0.07 & $<$3940$\pm$160      & $<1.2\pm0.3$    & 3400               & 0.5           \\
 BF\,Cyg    & M5            & 3355\,$\pm$\,75      & 3367                 & 1.35$\pm$0.06 & $<$0.24$\pm$0.01   & $>$1.23$\pm$0.07 & $<$3400$\pm$150      & $<0.2\pm0.3$    & 3400               & 0.0           \\
 CH\,Cyg    & M7            & 3100\,$\pm$\,80      & 3149                 & $\sim 1.6$    & $<$0.07$\pm$0.01   & $\sim 1.56$      & --                   & --              & 3100               & 0.0           \\
 QW\,Sge    & M5            & 3355\,$\pm$\,75      & 3367                 & 1.39$\pm$0.08 & $<$0.65$\pm$0.01   & $>$1.07$\pm$0.09 & $<$3730$\pm$190      & $<0.8\pm0.4$    & 3400               & 0.5           \\
 CI\,Cyg    & M5.5          & 3300\,$\pm$\,75      & 3312                 & 1.40$\pm$0.08 & $<$0.44$\pm$0.03   & $>$1.18$\pm$0.11 & $<$3500$\pm$240      & $<0.4\pm0.4$    & 3300               & 0.0           \\
 PU\,Vul    & M6.5$^b$      & 3170\,$\pm$\,75      & 3203                 & 1.38$\pm$0.07 & $<$0.29$\pm$0.01   & $>$1.24$\pm$0.08 & $<$3370$\pm$160      & $<0.2\pm0.3$    & 3200               & 0.0           \\
 V1329\,Cyg & M6.5$^b$      & 3170\,$\pm$\,75      & 3203                 & 1.46$\pm$0.06 & $<$0.35$\pm$0.01   & $>$1.29$\pm$0.07 & $<$3280$\pm$140      & $<$~$0.0\pm0.2$ & 3200               & 0.0           \\
 AG\,Peg    & M3.5$^b$      & 3510\,$\pm$\,75      & 3531                 & 1.19$\pm$0.06 & $<$0.08$\pm$0.01   & $>$1.15$\pm$0.07 & $<$3570$\pm$140      & $<0.5\pm0.3$    & 3500               & 0.5           \\
 Z\,And     & M4.5          & 3410\,$\pm$\,75      & 3421                 & 1.34$\pm$0.06 & $<$0.19$\pm$0.01   & $>$1.25$\pm$0.07 & $<$3360$\pm$140      & $<0.1\pm0.3$    & 3400               & 0.5           \\
  \hline
\end{tabular}
}}
\begin{list}{}{}
\item[{\bf References:}] spectral types are taken from
$^{[1]}$\citet{Mue1999}, total Galactic extinction adopted according to
$^{[6]}$\citet{Sch2011} and \cite{Sch1998}, IR from 2MASS
$^{[4]}$\citep{Phi2007} transformed to $^{[5]}$\citet{BeBr1988} photometric
system.
\item[{\bf Callibration by:}] $^{[2]}$\citet{Ric1999},
$^{[3]}$\citet{VBe1999}, $^{[7]}$\citet{Kuc2005}.
\item[{\bf $^{a}$}] finally adopted to chose the {\sl MARCS} model.
\item[{\bf $^{b}$}] slightly later spectral types were adopted than those by
\citet{Mue1999} based on spectra obtained close to inferior conjunctions
when they were not significantly influenced by nebulae \citep{Bel2000}.
\end{list}
\end{table*}

\begin{table*}
 \centering
  \caption{Estimated masses, radii, surface gravities, and orbital parameters of giants.}
\label{T-MaR}
  \begin{tabular}{@{}lllllll@{}}
  \hline
            & $M_{\rm{g}}$              & $R_{\rm{g}}$              & $\log{g}$              & $P_{\rm{orb}}$   & $K_{\rm{g}}$              & $i$                    \\
  Object    & [M$_{\sun}$]              & [R$_{\sun}$]              &                        & [day]            & [km\,s$^{-1}$]            & [$\degr$]              \\
  \hline
 EG\,And    & $1.1$--$2.4$\,$^{[1]}$    & $\sim 110$\,$^{a}$        & ~$0.4$\,--\,$0.7$      & 482.6\,$^{[1]}$  & 7.34$\pm$0.07\,$^{[1]}$   & 70\,$^{[11]}$          \\
 AX\,Per    & $\sim1.0$\,$^{[2,5]}$     & $132^{+22}_{-22}$\,$^{a}$ & ~$0.05$\,--\,$0.35$    & 682.1\,$^{[5]}$  & 7.81$\pm$0.21\,$^{[5]}$   & 70$\pm$3\,$^{[2]}$     \\
 T\,CrB     & $\sim0.9$\,$^{[3,4]}$     & $\sim 70$\,$^{[3,4]}$     & $\sim0.7$              & 227.6\,$^{[12]}$ & 23.89$\pm$0.17\,$^{[13]}$ & 60$\pm$5\,$^{[3]}$     \\
 FG\,Ser    & $1.7\pm0.7$\,$^{[5,6]}$   & $140^{+15}_{-13}$\,$^{a}$ & ~$0.05$\,--\,$0.6$     & 633.5\,$^{[5]}$  & 6.92$\pm$0.26\,$^{[5]}$   & 90\,$^{a}$             \\
 V443\,Her  & $\sim2.5$\,$^{[5,14]}$    & $166^{+27}_{-27}$\,$^{a}$ & ~$0.3$\,--\,$0.55$     & 599.4\,$^{[5]}$  & 2.52$\pm$0.21\,$^{[5]}$   & $\sim$30\,$^{[5,14]}$  \\
 V1413\,Aql & 1--2\,$^{b}$              & $147^{+70}_{-54}$\,$^{a}$ & $-0.2$\,--\,$0.8$      & 433.5\,$^{[15]}$ & --                        & --                     \\
 BF\,Cyg    & $\sim 2.2$\,$^a$          & $\leq240$\,$^{a}$         & $\geq 0.0$             & 757.2\,$^{[7]}$  & 6.72$\pm$0.24\,$^{[7]}$   & $\sim$70--90\,$^{[7]}$ \\
 CH\,Cyg    & $\sim2.0$\,$^{[9]}$       & $188 \pm 44$\,$^{[9]}$    & \,$0.0$\,--\,$0.4$     & 5689\,$^{[9]}$   & 4.45$\pm$0.12\,$^{[9]}$   & 84\,$^{[9]}$           \\
 QW\,Sge    & 1--2\,$^{b}$              & $156^{+42}_{-38}$\,$^{a}$ & $-0.15$\,--\,$0.6$     & 390.5\,$^{[16]}$ & --                        & --                     \\
 CI\,Cyg    & 0.85--1.27\,$^{[17]}$     & $197^{+37}_{-37}$\,$^{a}$ & $-0.35$\,--\,$0.15$    & 853.8\,$^{[13]}$ & 6.7$\pm$0.3\,$^{[13,17]}$ & 67--90\,$^{[18]}$      \\
 PU\,Vul    & $\sim0.8$\,$^{[10]}$      & $187 \pm 12$\,$^{[10]}$   & $\sim-0.2$             & 4901\,$^{[20]}$  & --                        & --                     \\
 V1329\,Cyg & 2.02$\pm$0.51\,$^{[7,8]}$ & $195 \pm 10$\,$^{a}$      & \,$0.0$\,--\,$0.3$     & 956.5\,$^{[19]}$ & 7.85$\pm$0.26\,$^{[7]}$   & 86$\pm$2\,$^{[19]}$    \\
 AG\,Peg    & $\geq 1.8$\,$^a$          & $137 \pm 11$\,$^{a}$      & $\geq 0.35$\,--\,$0.5$ & 818.2\,$^{[13]}$ & 5.44$\pm$0.20\,$^{[13]}$  & 90                     \\
 Z\,And     & $\sim$2\,$^{[5]}$         & $136^{+18}_{-19}$\,$^{a}$ & ~$0.4$\,--\,$0.6$      & 759.0\,$^{[5]}$  & 6.73$\pm$0.22\,$^{[5]}$   & $\sim 60$\,$^{a}$      \\
  \hline
\end{tabular}
\begin{list}{}{}
\item[{\bf References:}]
$^{[1]}$\citet{KeGa2016};  $^{[2]}$\citet{MiKe1992}; $^{[3]}$\citet{BeMi1998};  $^{[4]}$\citet{Sta2004};
$^{[5]}$\citet{Fek2000b};  $^{[6]}$\citet{Mue2000};  $^{[7]}$\citet{Fek2001};   $^{[8]}$\citet{Pri2003};
$^{[9]}$\citet{Hin2009};   $^{[10]}$\citet{Kat2012}; $^{[11]}$\citet{WiVa1997}; $^{[12]}$\citet{Kra1958};
$^{[13]}$\citet{Fek2000a}; $^{[14]}$\citet{Dob1993}; $^{[15]}$\citet{Poy2012};  $^{[16]}$\citet{MuJu2002};
$^{[17]}$\citet{Mik2006};  $^{[18]}$\citet{Ken1991}; $^{[19]}$\citet{ScSc1997}; $^{[20]}$\citet{Cun2018}.
\item[{\bf $^{a}$}] adopted -- see in the text (Appendix\,\ref{sc.npo}). 
\item[{\bf $^{b}$}] the mass of RGs is unknown and it was adopted a typical mass of giant in the S-type SySt 1--2 M$_{\sun}$ \citep{Mik2003}.
\end{list}
\end{table*}

\noindent expresses the upper bound to the RG radius ($R/a$) to permit the
system to remain detached.  The known limit on the mass ratio enables us to
estimate the minimal mass of the RG ($M_{\rm{g}}$) through the known mass
function and thus a lower limit to the surface gravity $\log{g}$
(Table\,\ref{T-elogg} -- column ({\sl{I}})) when the radius of the star is
estimated from the known orbital period ($P_{\rm{orb}}$), observed
rotational velocity ($V_{\rm{rot}} \sin{i}$), and inclination ($i$), and
when the synchronous rotation with orbital motion is adopted.  The obtained
values can be verified using the aforementioned \citet{Kuc2005} calibration
binding $T_{\rm{eff}}$, $\log{g}$, and colour in a linear relation. 
However, a limitation of this approach is that the calibration is based on
normal RGs, while symbiotic giants at least in some cases could turn
out to be somewhat brighter luminosity class.  Our previous results indicate
that parameters of at least yellow SySt may need to be revised
\citep{Gal2017}.  It is further supported by ellipsoidal light curves being
discovered in more and more systems including roughly half of systems
studied here \citep[see discussion in][and Appendix \ref{sc.npo} for
individual cases]{Mik2012} and the new Gaia distances which indicated that
giants in many of SySt must have significantly larger diameters, be
significantly colder, and of higher luminosity class -- bright giants rather
than a normal giants.  Furthermore, as we used the total Galactic extinction
to de-redden IR colours, the obtained values (Table\,\ref{T-est} and
Table\,\ref{T-elogg} -- column ({\sl{III}})) should be treated as the upper
limit to $\log{g}$ at least in the cases when the adopted $E$($B-V$) is
large.  The final adopted values of surface gravities are listed in the
right-most column of Table\,\ref{T-elogg} and are repeated in
Table\,\ref{T-est}.

Additionally, the spectra contain wavelength shifts from radial velocity
doppler effects and line broadening from stellar rotation as well as
small-scale micro- ($\xi_{\rm t}$) and macro- ($\zeta_{\rm t}$) turbulence. 
$\xi_{\rm t}$ and $\zeta_{\rm t}$ were set in our calculations to typical
values for cool Galactic RGs, 2 and 3\,km\,s$^{-1}$,
respectively.  Corrections for the velocity shifts (contributed mainly by
radial velocities) were performed with the use of the values obtained
through the fit of the synthetic spectra to the observed ones, during the
initial phase of estimating the input parameters, which were then set as the
fixed values during subsequent more accurate calculations of the chemical
composition.  In our spectra of giant stars with relatively narrow
absorption features the largest contribution to the broadening of the
spectral lines comes from the rotational velocity of the star ($V_{\rm{rot}}
\sin{i}$).  We measured these values by adjusting them as free parameters in
the course of the fitting with the synthetic spectra.  In the case of the
$K$-band spectra, we were also able to measure values of the rotational
velocity directly from the full width at half-maximum (FWHM) of the six
relatively strong unblended atomic lines (\ion{Ti}{i}, \ion{Fe}{i},
\ion{Sc}{i}).  The values obtained with both methods are generally in very
good agreement (Table\,\ref{T-js}), except the case of AG\,Peg, in which the
value derived from the synthetic fit is somewhat overestimated, due to the
influence of an additional absorption component present in the line profiles
(see Section\,\ref{sc.AaR} for more details).

\section{Analysis and results} \label{sc.AaR}

Elemental abundances were measured using the spectrum synthesis code
\citep[{\sl WIDMO};][]{Sch2006} which applies local thermodynamic
equilibrium analysis based on 1D hydrostatic model atmospheres
\citep[{\sl MARCS};][]{Gus2008}.  All details of our methods are described
by \citet{Gal2016, Gal2017}.  The lists of the atomic and molecular lines
with the excitation potentials and $gf$-values for transitions are the same
as in our previous studies of symbiotic giants.  For the $H$-band region,
the atomic data are from the list by \citet{MeBa1999}.  For $K$- and $K{\rm
r}$-band regions the lists from the Vienna Atomic Line Data base
\citep{Kup1999} were used.  For the molecular data, we used line lists by
\citet{Goo1994} for CO, by \citet{Kur1999} for OH, and by \citet{Sne2014}
for CN.

In summary, the abundance calculations were performed as follows.  We
sampled a number of {\sl MARCS} models around those with the adopted best
values of $T_{\rm{eff}}$, $\log{g}$, and $\xi$, testing various
metallicities and differently CN-cycled models as reflected in the carbon
isotopic ratio $^{12}$C/$^{13}$C.  The simplex algorithm \citep{Bra1998} was
used for $\chi^2$ minimization in the space of 9 free parameters: 7 for
abundances of C, N, O, Sc, Ti, Fe, and Ni, and 2 for rotational velocities
from the $H$- and $K$-band region spectra.  In most cases, when the $K_{\rm
r}$-band spectra were available there we performed additional fits to this
spectral region with abundances of carbon isotopes $^{12}$C and $^{13}$C,
and rotational velocity as free parameters.  The above procedure was
repeated for each target to choose the model with the best-matching
metallicity.  The final abundances derived on the scale of $\log{\varepsilon
(X)} = \log{(N(X)N(H)^{-1})} + 12.0$, are summarized in
Table\,\ref{T-finAbu} together with the ratio of carbon isotopes,
$^{12}$C/$^{13}$C, and corresponding uncertainties.  The observed spectra
contain some artifacts -- the lines that have no counterparts in the line
lists or in the synthetic spectrum that have bad data for atomic
transitions.  Suitable masks were prepared to exclude these lines from the
analysis.  Synthetic fits to the observed spectra of EG\,And are shown in
Figs\,\ref{F1}--\ref{F3} and visualizations of the fits to all the observed
spectra are available in the online Appendix\,\ref{AppS}.

Our formal fitting errors were typically several hundredths of dex but up to
nearly $\sim$0.2\,dex in a few cases of abundance values for Ti, Sc, and Ni. 
Nevertheless, the actual uncertainties of the chemical composition are
higher and come mainly from uncertainties in stellar parameters, to which we
assign uncertainties of $\sim$100\,K in effective temperature, 0.5\,dex in
$\log{g}$, and $\sim$0.25\,km\,s$^{-1}$ in the case of $\xi_{\rm{t}}$.  To
examine how significantly this reflects in the accuracy of determining
abundances we executed additional calculations, varying the atmospheric
parameters by these values.  The resulting values are shown in
Table\,\ref{TA-AbuSens} in the online Appendix\,\ref{AppT}.

\begin{figure*}
   \centering
   \includegraphics[width=\hsize]{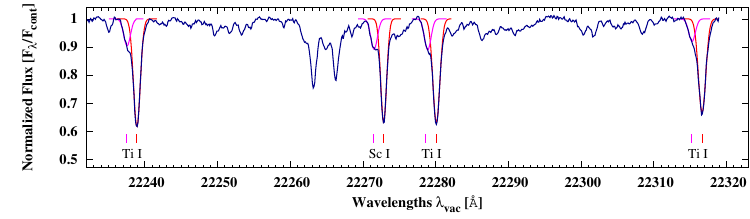}
      \caption{The spectrum of AG\,Peg in $K$-band region (blue line).  The
components of the spectral line profiles of 4 isolated, minimally blended
spectral features from Ti and Sc (Table\,\ref{T-4AGPeg}) are marked with red
Gaussians for the stellar component originating from the RG, and
with a magenta line for the additional, blue-shifted component, which must
be a manifestation of some kind of interactions in the system (see the text
for the details).}
         \label{F4}
   \end{figure*}

In the case of AG\,Peg we found an additional source of error in the
measured abundances.  An additional component is present in the line
profiles, clearly visible in the isolated and not significantly blended,
strong titanium and scandium lines in the $K$-band region (Fig.\,\ref{F4}). 
The presence of this additional component can be seen also in the residuals
of the observed and the synthetic spectra of the $H$- and $K_{\rm r}$-band
regions (see Figs\,\ref{FC27} and \ref{FC29} in the Appendix\,\ref{AppS}),
but line blending is too strong for reliable measurements.  We measured the
positions of these spectral features in the $K$-band region and compared
them to the laboratory wavelengths, to obtain the radial velocities of the
components.  The obtained values can be found in Table\,\ref{T-4AGPeg}.  The
radial velocity of the main component (the RG) is V$_{\rm Rad}$\,(A) =
$-19.6 \pm 1.1$ km\,s$^{-1}$.  The additional, secondary component is
shifted in relation to this stellar by Shift\,(B)=$-17.7 \pm 1.1$ towards
the blue, which corresponds to its radial velocity V$_{\rm Rad}$\,(B) =
$-37.3 \pm 2.2$ km/s.  We measured the equivalent widths (EW) of both
components (Table\,\ref{T-4AGPeg}) from the originally observed spectrum
(EW$_{\rm O}$\,(A) and EW$_{\rm O}$\,(B)) as well as from the synthetic
spectrum (EW$_{\rm S}$\,(A)) and from the residuals (EW$_{\rm S}$\,(B)). 
Using these values we estimated that the abundances derived for AG\,Peg can
be overestimated (i.e., in reality, they should be smaller) as a result of
contribution from the secondary component to the measured profiles by $\sim
0.02$\,dex.  However, this value is smaller than the formal errors of fit
(see Table\,\ref{T-finAbu}).

%

   \begin{figure}
   \centering
   \includegraphics[width=\hsize]{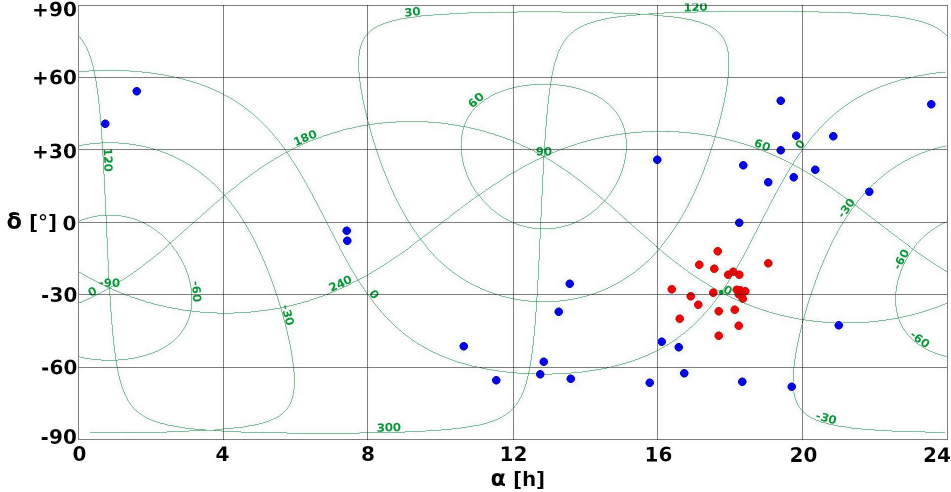}
      \caption{The position of southern \citep{Gal2016,Gal2017} and northern
(this work) SySt in the equatorial coordinate system representing objects
from the bulge (red points) and other systems generally from the Galactic
disc (blue points).  A superimposed green grid presents the Galactic
coordinate system.}
         \label{F5}
   \end{figure}
%

A few of our targets have previously measured abundances.  For CH\,Cyg
\citet{Sch2006} employed wide ranges of the $H$ and $K$-band regions.  The
resulting values agree with the current ones (derived only from the short
$H$-band spectrum) within the error borders.  EG\,And and AX\,Per have
counterparts within 1 arcsec with their J2000 coordinates in the data base of
Apache Point Observatory Galactic Evolution Experiment \citep[{\sl
APOGEE};][]{Maj2017, Jon2020}.  The abundances agree with our estimates with
exception of nitrogen -- where {\sl APOGEE} values are lower by
$\sim$0.3\,dex.

\begin{table*}
  \caption{The final abundances derived on the scale of $\log{\epsilon}(X) =
\log{(N(X) N(H)^{-1})} + 12.0$, relative to the Solar abundances, carbon
$^{12}$C/$^{13}$C isotopic ratio, and uncertainties$\,^a$.}
  \label{T-finAbu}
  \begin{tabular}{@{}lrrrrrrrr@{}}
  \hline
  \hline
                    & C                   & N         & O              & Sc$^b$         & Ti             & Fe             & Ni             &  $^{12}$C/$^{13}$C   \\
                    & $\log{\epsilon(X)}$ &           &                &                &                &                &                &                      \\
                    & [$X$]$^c$           &           &                &                &                &                &                &                      \\
  \hline
EG\,And             & $ 7.70\pm0.03$ & $ 7.81\pm0.04$ & $ 8.37\pm0.01$ & $ 3.34\pm0.05$ & $ 4.80\pm0.04$ & $ 6.93\pm0.01$ & $ 5.90\pm0.07$ & ~7.0$\pm$0.3         \\
                    & $-0.73\pm0.08$ & $-0.02\pm0.09$ & $-0.32\pm0.06$ & $+0.18\pm0.09$ & $-0.13\pm0.08$ & $-0.54\pm0.05$ & $-0.30\pm0.11$ &                      \\
AX\,Per             & $ 7.84\pm0.01$ & $ 8.05\pm0.03$ & $ 8.41\pm0.02$ & $ 3.87\pm0.07$ & $ 5.04\pm0.06$ & $ 7.21\pm0.06$ & $ 6.26\pm0.06$ & ~9.5$\pm$0.3         \\
                    & $-0.59\pm0.06$ & $+0.22\pm0.08$ & $-0.28\pm0.07$ & $+0.71\pm0.11$ & $+0.11\pm0.10$ & $-0.26\pm0.10$ & $+0.06\pm0.10$ &                      \\
T\,CrB              & $ 8.40\pm0.02$ & $ 8.65\pm0.04$ & $ 8.79\pm0.01$ & ...            & $ 5.12\pm0.09$ & $ 7.82\pm0.04$ & $ 6.57\pm0.06$ & ...                  \\
                    & $-0.03\pm0.07$ & $+0.82\pm0.09$ & $+0.10\pm0.06$ & ...            & $+0.19\pm0.13$ & $+0.35\pm0.08$ & $+0.37\pm0.10$ &                      \\
FG\,Ser             & $ 8.08\pm0.01$ & $ 7.83\pm0.03$ & $ 8.52\pm0.01$ & ...            & $ 4.79\pm0.06$ & $ 7.39\pm0.02$ & $ 6.23\pm0.05$ & ...                  \\
                    & $-0.35\pm0.06$ & ~$0.00\pm0.08$ & $-0.17\pm0.06$ & ...            & $-0.14\pm0.10$ & $-0.08\pm0.06$ & $+0.03\pm0.09$ &                      \\
V443\,Her           & $ 8.18\pm0.02$ & $ 8.07\pm0.03$ & $ 8.62\pm0.01$ & ...            & $ 4.97\pm0.10$ & $ 7.45\pm0.04$ & $ 6.29\pm0.05$ & ...                  \\
                    & $-0.25\pm0.07$ & $+0.24\pm0.08$ & $-0.07\pm0.06$ & ...            & $+0.04\pm0.14$ & $-0.02\pm0.08$ & $+0.09\pm0.09$ &                      \\
V1413\,Aql          & $ 8.10\pm0.05$ & $ 7.74\pm0.10$ & $ 8.31\pm0.03$ & ...            & $ 4.45\pm0.14$ & $ 7.35\pm0.07$ & $ 6.35\pm0.12$ & ...                  \\
                    & $-0.33\pm0.10$ & $-0.09\pm0.15$ & $-0.38\pm0.08$ & ...            & $-0.48\pm0.18$ & $-0.12\pm0.11$ & $+0.15\pm0.16$ &                      \\
BF\,Cyg             & $ 7.87\pm0.03$ & $ 8.23\pm0.08$ & $ 8.52\pm0.01$ & $ 3.89\pm0.15$ & $ 4.89\pm0.10$ & $ 7.22\pm0.03$ & $ 6.02\pm0.06$ & ~6.1$\pm$0.5         \\
                    & $-0.56\pm0.08$ & $+0.40\pm0.13$ & $-0.17\pm0.06$ & $+0.73\pm0.19$ & $-0.04\pm0.14$ & $-0.25\pm0.07$ & $-0.18\pm0.10$ &                      \\
CH\,Cyg             & $ 8.26\pm0.01$ & $ 8.20\pm0.02$ & $ 8.66\pm0.01$ & ...            & $ 5.06\pm0.08$ & $ 7.60\pm0.05$ & $ 6.39\pm0.07$ & ...                  \\
                    & $-0.17\pm0.06$ & $+0.37\pm0.07$ & $-0.03\pm0.06$ & ...            & $+0.13\pm0.12$ & $+0.13\pm0.09$ & $+0.19\pm0.11$ &                      \\
QW\,Sge             & $ 8.30\pm0.03$ & $ 8.20\pm0.07$ & $ 8.67\pm0.02$ & $ 4.25\pm0.12$ & $ 5.28\pm0.09$ & $ 7.57\pm0.10$ & $ 6.54\pm0.10$ & 13.9$\pm$0.8         \\
                    & $-0.13\pm0.08$ & $+0.37\pm0.12$ & $-0.02\pm0.07$ & $+1.09\pm0.16$ & $+0.35\pm0.13$ & $+0.10\pm0.14$ & $+0.34\pm0.14$ &                      \\
CI\,Cyg             & $ 7.97\pm0.04$ & $ 8.17\pm0.07$ & $ 8.50\pm0.02$ & $ 4.52\pm0.14$ & $ 5.25\pm0.06$ & $ 7.37\pm0.03$ & $ 6.17\pm0.10$ & 12.6$\pm$1.1         \\
                    & $-0.46\pm0.09$ & $+0.34\pm0.12$ & $-0.19\pm0.07$ & $+1.36\pm0.18$ & $+0.32\pm0.10$ & $-0.10\pm0.07$ & $-0.03\pm0.14$ &                      \\
PU\,Vul             & $ 8.00\pm0.02$ & $ 7.97\pm0.03$ & $ 8.34\pm0.01$ & $ 3.37\pm0.09$ & $ 4.35\pm0.06$ & $ 7.10\pm0.02$ & $ 5.90\pm0.09$ & 16.2$\pm$0.8         \\
                    & $-0.43\pm0.07$ & $+0.14\pm0.08$ & $-0.35\pm0.06$ & $+0.21\pm0.13$ & $-0.58\pm0.10$ & $-0.37\pm0.06$ & $-0.30\pm0.13$ &                      \\
V1329\,Cyg          & $ 8.45\pm0.03$ & $ 8.27\pm0.07$ & $ 8.66\pm0.02$ & $ 4.36\pm0.08$ & $ 5.09\pm0.06$ & $ 7.59\pm0.05$ & $ 6.35\pm0.06$ & 24.0$\pm$1.5         \\
                    & $+0.02\pm0.08$ & $+0.44\pm0.12$ & $-0.03\pm0.07$ & $+1.20\pm0.12$ & $+0.16\pm0.10$ & $+0.12\pm0.09$ & $+0.15\pm0.10$ &                      \\
AG\,Peg             & $ 7.62\pm0.03$ & $ 7.82\pm0.06$ & $ 8.18\pm0.02$ & $ 3.60\pm0.04$ & $ 4.61\pm0.05$ & $ 6.96\pm0.02$ & $ 5.81\pm0.03$ & ~5.2$\pm$0.1         \\
                    & $-0.81\pm0.08$ & $-0.01\pm0.11$ & $-0.51\pm0.07$ & $+0.44\pm0.08$ & $-0.32\pm0.09$ & $-0.51\pm0.06$ & $-0.39\pm0.07$ &                      \\
Z\,And              & $ 8.11\pm0.03$ & $ 8.17\pm0.06$ & $ 8.56\pm0.02$ & $ 4.13\pm0.12$ & $ 5.01\pm0.11$ & $ 7.41\pm0.04$ & $ 6.33\pm0.11$ & 10.5$\pm$0.9         \\
                    & $-0.32\pm0.08$ & $+0.34\pm0.11$ & $-0.13\pm0.07$ & $+0.97\pm0.16$ & $+0.08\pm0.15$ & $-0.06\pm0.08$ & $+0.13\pm0.15$ &                      \\
  \hline
Sun                 & $ 8.43\pm0.05$ & $ 7.83\pm0.05$ & $ 8.69\pm0.05$ & $ 3.16\pm0.04$ & $ 4.93\pm0.04$ & $ 7.47\pm0.04$ & $ 6.20\pm0.04$ &                      \\
  \hline
  \hline
\end{tabular}
\begin{list}{}{}
\small{
\item[$^a$] 3$\sigma$.
\item[$^b$] The abundance of scandium is based on only one strong Sc\ion{I}
\,line at $\lambda \sim$ 22 272.8\,\AA\/ and it may be less reliable than
other abundances.  Broadening of the IR scandium lines by hyperfine
structure has not been included in the analysis \citep[see][]{Mik2014}.
\item[$^c$] Relative to the Sun [$X$] abundances in respect to the solar
composition of \citet{Asp2009} and \citet{Sco2015}.
}
\end{list}
\end{table*}

\begin{table*}
  \caption{The measured radial velocities and EWs of the components of the spectral line profiles in the $K$-band spectrum of AG\,Peg.}
  \label{T-4AGPeg}
  \begin{tabular}{@{}llllllllll@{}}
  \hline
  \hline

Line        & $\lambda_{\rm{Lab. (vac)}}$ & $\lambda_{\rm{Lab. (air)}}$ & $\lambda_{\rm{Meas. (air)}}$ & V$_{\rm Rad}$\,(A) & Shift (B) & EW$_{\rm O}$\,(A) & EW$_{\rm O}$\,(B) & EW$_{\rm S}$\,(A) & EW$_{\rm S}$\,(B) \\
            & [\AA]                       & [\AA]                       & [\AA]                        & [km/s]             & [km/s]    & [m\AA]            & [m\AA]            & [m\AA]            & [m\AA]            \\
\hline
\ion{Ti}{1} & 22238.90                    & 22232.84                    & 22231.40                     & -19.42             & -18.07    & 448.6             & 111.8             & 467.2             & 98.0              \\
\ion{Sc}{1} & 22272.81                    & 22266.73                    & 22265.32                     & -18.99             & -18.58    & 403.5             & 110.7             & 451.7             & 105.6             \\
\ion{Ti}{1} & 22280.11                    & 22274.01                    & 22272.57                     & -19.38             & -16.96    & 432.3             & 111.8             & 473.7             & 106.5             \\
\ion{Ti}{1} & 22316.67                    & 22310.62                    & 22309.09                     & -20.56             & -17.33    & 411.6             & 109.2             & 453.1             & 105.5             \\
  \hline
  \hline
\end{tabular}
\end{table*}

   \begin{figure}
   \centering
   \includegraphics[width=\hsize]{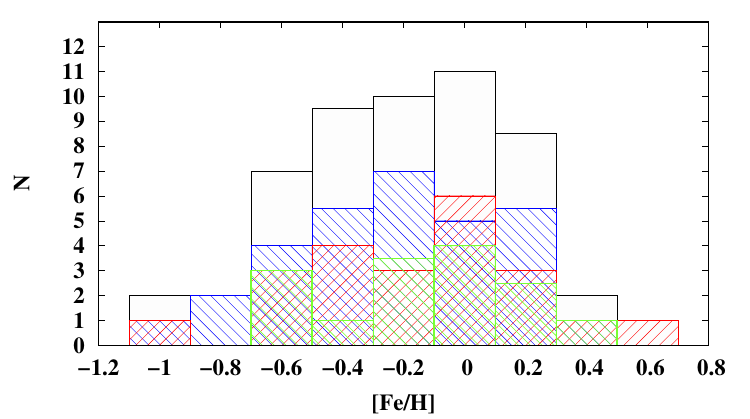}
      \caption{The distribution for the number 'N' of objects, counted at
0.2\,dex intervals, as a function of metallicity ([Fe/H]) for all symbiotic
giants studied by us so far (black).  The whole sample is divided into two
subsamples: bulge stars (red) and other objects mostly from the Galactic
disc (blue).  The northern sample measured in this paper is shown with
green.}
         \label{F6}
   \end{figure}
%

   \begin{figure*}
   \centering
   \includegraphics[width=\hsize]{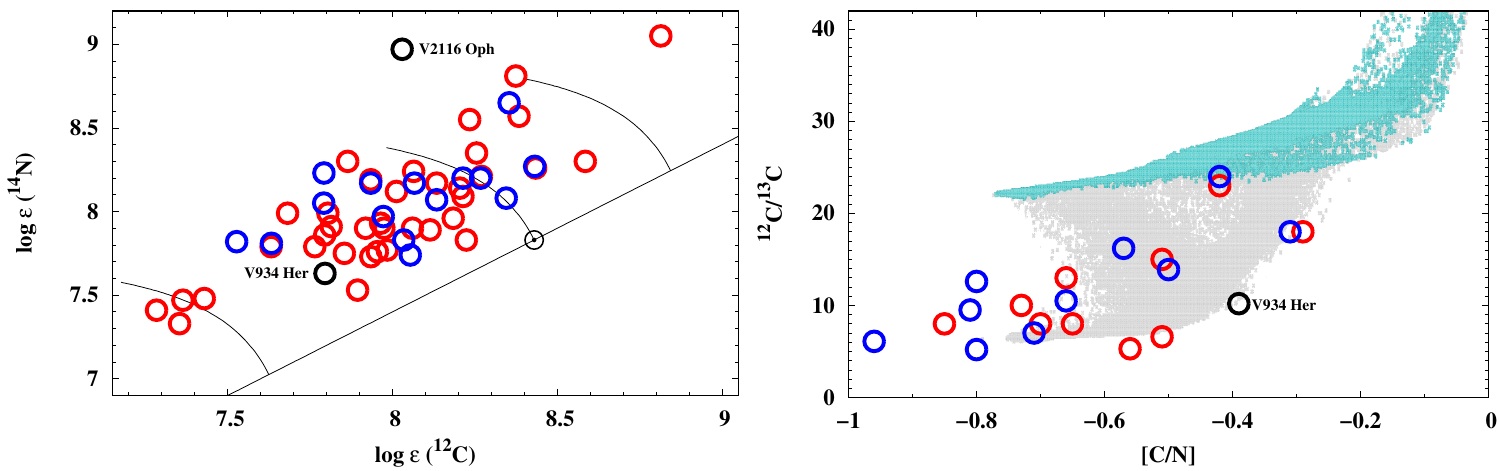}
      \caption{{\sl Left:} Nitrogen versus carbon for the SySt from northern
(blue) and southern (red) samples.  Black circles mark two SySt with an
accreting neutron star (V934\,Her and V2116\,Oph).  The solid line
represents scaled solar abundances, [$^{12}$C/Fe] = 0 and [$^{14}$N/Fe] = 0
whereas the solid curves delineate constant $^{12}$C + $^{14}$N.  {\sl
Right:} $^{12}$C/ $^{13}$C versus [C/N] from our study compared with
theoretical model predictions for synthetic populations computed with the
{\sl BGM} \citep{Lag2017} with the effects of thermohaline instability (grey dots)
and without (turquoise dots).}
         \label{F7}
   \end{figure*}
%

\section{Concluding discussion} \label{sc.dis}

We have derived here the 'photospheric' chemical abundances (C, N, O and
elements around the iron peak: Sc, Ti, Fe, and Ni) for a sample of 14 RGs
in the classical S-type SySt.  They are located in the northern
hemisphere with the one exception of FG\,Ser which is located almost exactly
at the celestial equator.  While the southern sample \citep{Gal2016,Gal2017}
is dominated by objects concentrated around the Galactic centre, the
northern sample is dominated by the Galactic disc and halo
(Fig.\,\ref{F5}) and will make a useful complement to the southern sample
in our ongoing comprehensive analyses of the kinematics and chemical
evolution of SySt in the Galactic populations (Galan et al. 
-- in preparation).

Metallicity is one of the most important parameters in the study of
evolution in Galactic stellar populations as well as the history of
interactions in binary systems, as the metallicity impacts the efficiency of
the mass loss from evolved giant components, and thus on the rate of the
mass exchange.  As an indicator of metallicity, we use the abundance of iron
that is often treated as its proxy.  The distribution of the number of
objects as a function of [Fe/H] is shown in Fig.\,\ref{F6}.  The global
distribution for all currently studied SySt has a maximum at a solar value
[Fe/H] = 0.0\,dex.  It is somewhat asymmetrical and shifted towards
sub-solar values with a median distribution at $-0.2$\,dex.  Most of
the symbiotic giants are characterized by slightly sub-solar metallicity, as
expected for an older disc population.

As with all previously analyzed symbiotic giants, those from the current
northern sample have measured abundances of C, N, and O similar to
those in single Galactic M giants with enhanced $^{14}$N, depleted $^{12}$C,
and decreased $^{12}$C/$^{13}$C \citep[see][]{Gal2016, Gal2017}.  The
abundances of $^{14}$N versus $^{12}$C for the sample of all SySts which we
have studied so far are shown in Fig.\,\ref{F7} (left), and they prove
that all these objects have experienced the 1-st dredge-up.  This is also
confirmed by the low $^{12}$C/$^{13}$C, which, however, being too low with
respect to the theoretical predictions \citep{Lu2008} suggests
that the mixing resulting from the 1-st dredge-up is insufficient to explain
the observed abundances of C, so some additional processes must be
occurring.  The phenomenon of thermohaline mixing \citep[see eg.][]{CaZ2007}
is a likely possibility.

\citet{Lag2019} presented the first comparison between the synthetic
populations computed with the improved Besan\c{c}on Galaxy model \citep[{\sl
BGM};][]{Lag2017} and the C and N abundances derived by the Gaia-ESO survey
in field stars and in open and globular clusters.  They showed that
additional mixing has an impact on the C and N abundances as well as on
$^{12}$C/$^{13}$C, which turned out to be an even better parameter for
constraining extra mixing on the RGB, and works efficiently in lower mass
and older giant stars (M $\leqslant$ 2.2\,M$_{\sun}$), especially at lower
metallicities.  Fig.\,\ref{F7}--right shows our sample of symbiotic giants
in the $^{12}$C/$^{13}$C vs [C/N] plane compared to two models -- the first
with and the second without the thermohaline instability taken into account. 
Our results confirm that symbiotic giants may have gone through this phase
of mixing with thermohaline instability. It seems that binary
interaction has not significantly affected the evolution of symbiotic
giants that are similar to normal M giants.

   \begin{figure}
   \centering
   \includegraphics[width=\hsize]{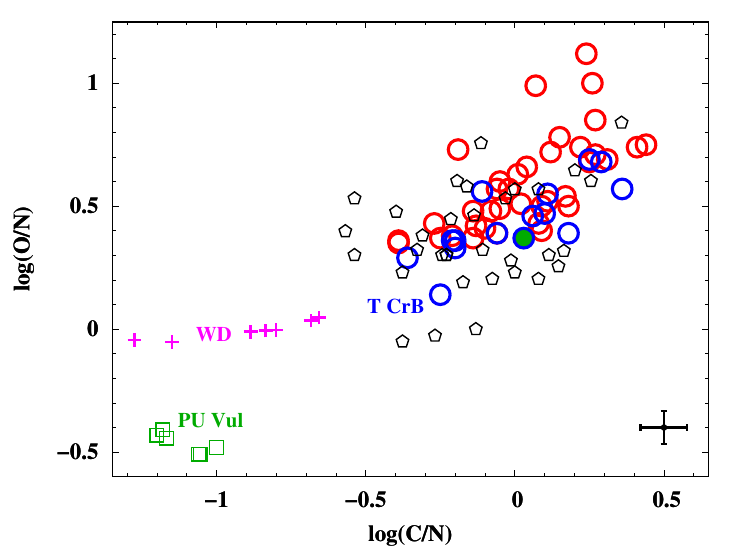}
      \caption{O/N versus C/N from the photospheric abundances.  Northern
(blue circles -- this work) and southern \citep[red circles
--][]{Gal2016, Gal2017} samples are compared with values from nebular lines
\citep[pentagons --][]{Nus1988, ScSc1990, Per1995, Sch2006}.  The
position of PU\,Vul from our measurements of photospheric abundances is
highlighted by filling the symbol with green colour. Squares show PU\,Vul
during outburst \citep{Vog1992}.  Crosses represent theoretical predictions
for nova ejecta from CO WD with 0.65\,M$_{\sun}$ \citep{KoPr1997}.  In the
bottom-right corner is shown typical $3\sigma$ error bar.}
         \label{F10}
   \end{figure}
%

The ratios of the photospheric abundances O/N and C/N are shown in
Fig.\,\ref{F10}.  Both our northern and southern samples show C/N $<$
O/N, which compared to the theoretical values obtained by \citet{Lu2008}
suggests that the cool components of our symbiotic systems are low-mass
giants ($M < 2.5$\,M$_{\sun}$) that have not yet undergone or have undergone
only an inefficient 3-rd dredge-up.  

\citet{Gal2017} have shown that these ratios obtained from nebular emission
lines are shifted somewhat towards lower O/N and C/N ratios compared to
values from 'photospheric' abundances, which may be due to pollution of the
nebula by hot component outburst ejecta.  This shift is shown on example of
PU\,Vul.  The photospheric abundances of PU\,Vul locate it among the other
symbiotic giants, whereas those derived for the nebula move it towards to
abundances predicted for the nova ejecta indicating the WD mass of
0.6 M$_{\sun}$ \citep{Kat2012b}.  Since modelling of the outburst with the
predictions for the nova ejecta from 0.65 M$_{\sun}$ CO WD \citep{KoPr1997},
the behaviour of PU\,Vul in Fig.\,\ref{F10} indicates that its abundances
are affected by the nova ejecta.

\begin{figure}
   \centering
   \includegraphics[width=\hsize]{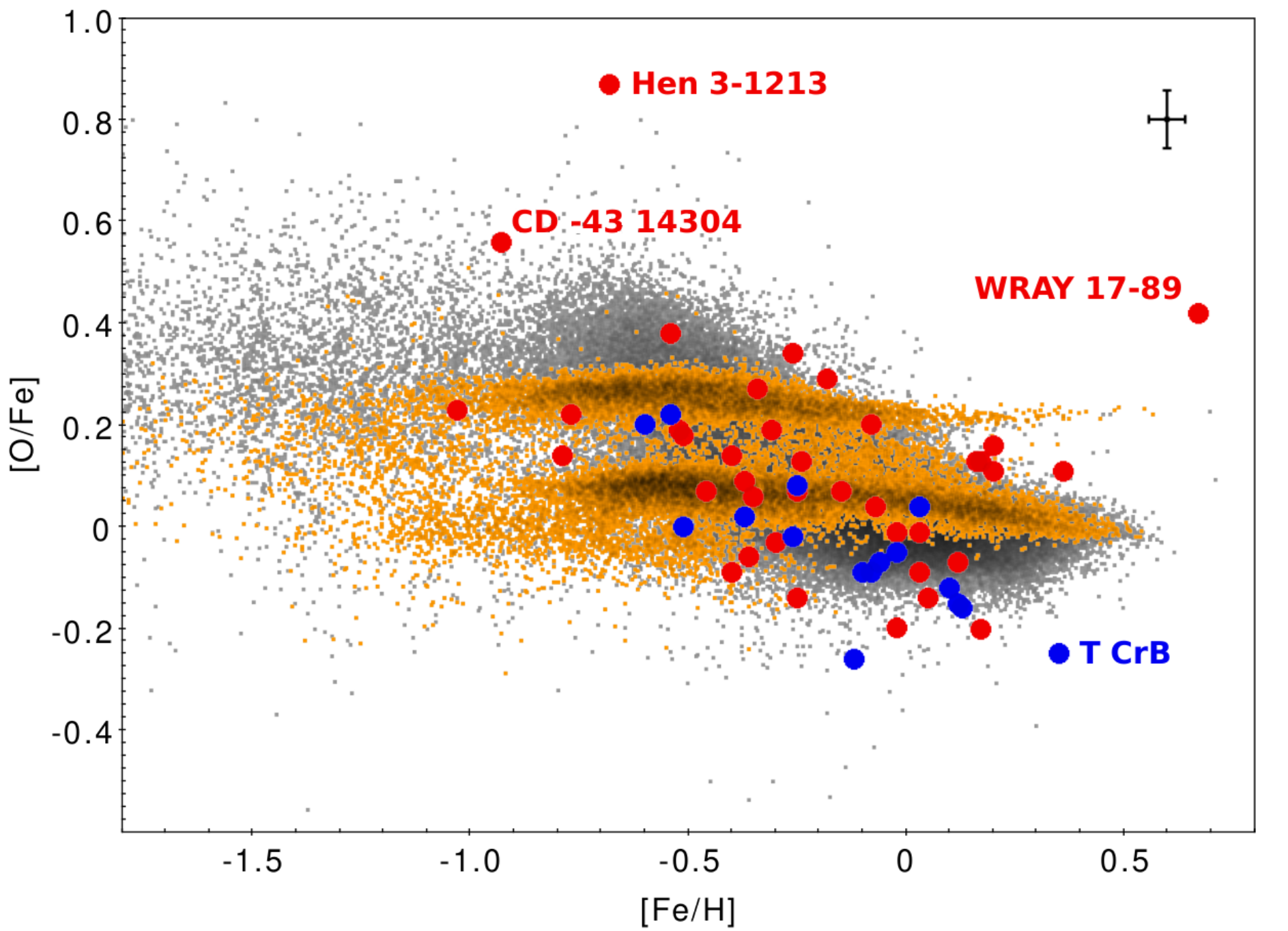}
      \caption{Oxygen relative to iron for our northern (blue circles) and
southern (red circles) samples compared to the abundances coming from the
entire selected sample of the {\sl APOGEE} DR16 release (gray points) and
the extracted sample of giant stars (orange points) corresponding to the
atmospheric parameters (3100 $\leq T_{\rm{eff}} \leq$ 4100\,K, and 0 $\leq
\log{g} \leq$ 1.5) similar to our sample of SySt.  The {\sl APOGEE}
abundances have been scaled to the solar composition by \citet{Asp2009} and
\citet{Sco2015} for oxygen and iron, respectively. Typical $3\sigma$
error bar is shown in the top-right corner.}
         \label{F8}
   \end{figure}

In addition to Fe providing information about metallicity, we also measured
the abundances of two $\alpha$ elements: O and Ti.  Due to the well-known
fact that there are differences in the lifetimes of objects delivering
these elements to the interstellar medium, they are especially useful in
studying the chemical evolution of galaxies and the formation of stellar
populations.  The relative abundances of O and Ti in relation to Fe for our
northern and southern samples of SySts are shown in Fig.\,\ref{F8}
where they are compared with results from the {\sl APOGEE} project.  The
data of the DR16 release \citep{Jon2020} were filtered to reject the
measurements with excessive uncertainty and various defects according
to the criteria: we used only the data from the spectra with S/N > 70, and
all those with flags\footnote{see the web page
https://www.sdss.org/dr16/irspec/ of the {\sl APOGEE} project with a
description of how to use the data} {\sl STARFLAG} and {\sl ASPCAPFLAG}
equal to zero have been rejected.  Next, we removed the potential binary
systems (about 6000 objects) from the sample using the
following criterion: $(VSCATTER > 1)$ and $((VSCATTER / VERR_MED) > 5)$.

Fig.\,\ref{F8} shows [O/Fe] vs [Fe/H] diagram for whole our sample of SySt
compared to {\sl APOGEE} results for a subsample of giants characterized by
atmospheric parameters ($T_{\rm{eff}}$: 3100 -- 4100\,K, $\log{g}$: 0.0 --
1.5), similar to our sample of SySts.  It is notable that this sample of
giants splits into two clearly separated 'sequences' which represent an
$\alpha$-enhanced old population and the less $\alpha$-enhanced
young population at lower [O/Fe].  Our objects coincide well with giants
from {\sl APOGEE} taking into account the uncertainty (see
Tables\,\ref{T-finAbu} and \ref{TA-AbuSens}) with the exception of
WRAY\,17-89 and Hen\,3-1213.  At least in some cases, discrepancies can
result from an incorrect effective temperature of the giant.  This can occur
easily in a SySt because of the strong contribution from
the nebular continuum.  An example is CD$-43\degr14304$ \citep{Gal2017}. 
The metallicity ([Fe/H]) of symbiotic giants ranges from $\sim -1$ to
$+0.6$\,dex, in most cases being consistent with membership in the
disc/thick disc population, or in some cases the extended thick disc or
halo.

%

\begin{figure}
   \centering
   \includegraphics[width=\hsize]{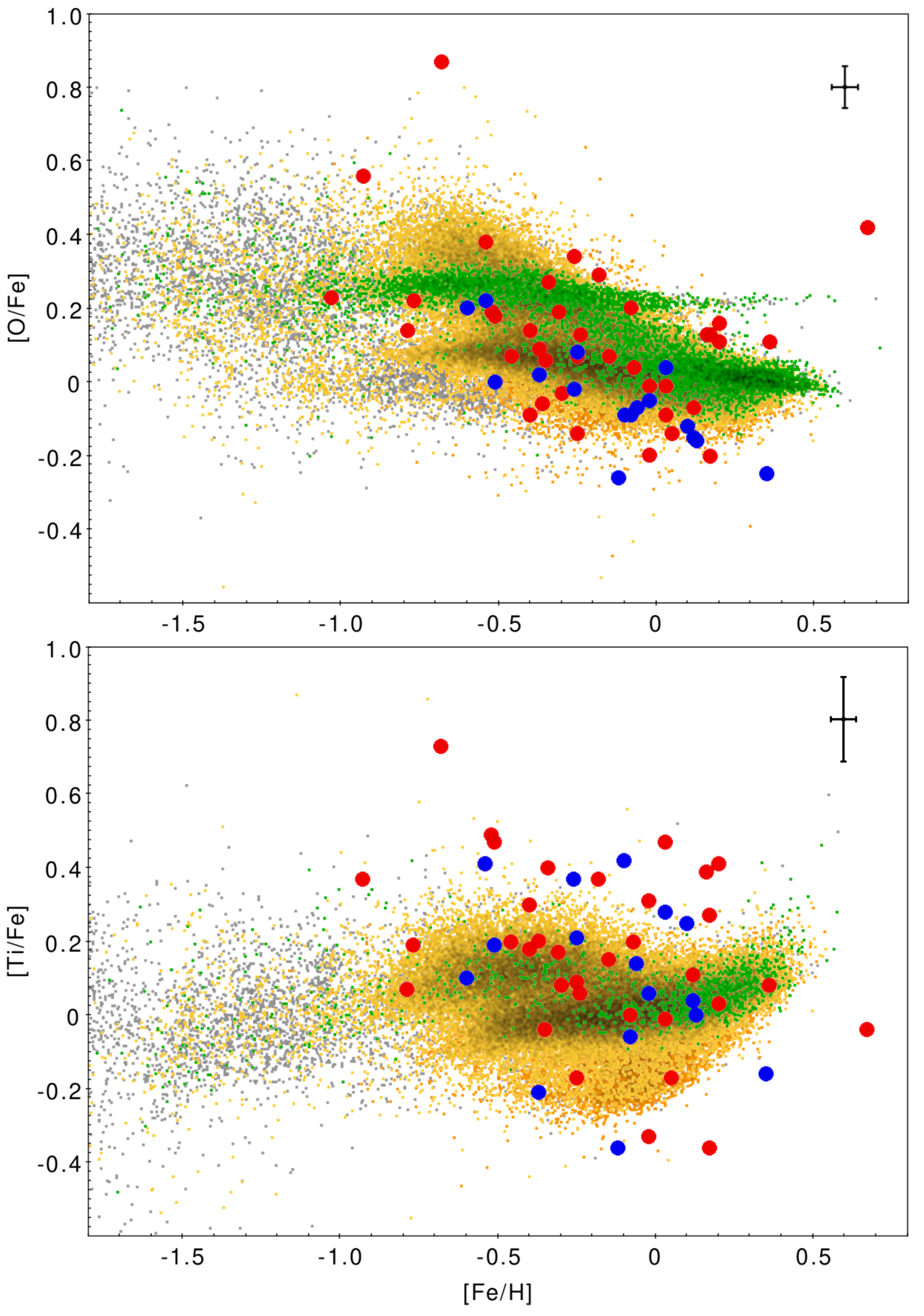}
      \caption{[O/Fe] vs [Fe/H] ({\sl Top}) and [Ti/Fe] vs [Fe/H] ({\sl
Bottom}) for our northern (blue circles) and southern (red circles)
samples, compared with various stellar populations as extracted from
the {\sl APOGEE} data: {\sl Halo} with gray points, {\sl Thin-} and {\sl
Thick-discs} with orange and yellow, respectively and {\sl Bulge} with
green.  Thin- and Thick-discs samples percolate significantly in this
representation, with the latter characterized by slightly lower [O/Fe] and
[Ti/Fe] values and hidden under the set of points from the former. 
Typical $3\sigma$ error bars are shown in the top-right corners.}
         \label{F9}
   \end{figure}

In Fig.\,\ref{F9}, the relative abundances of [O/Fe] and [Ti/Fe] versus
[Fe/H] of our SySts are presented in comparison to stars from various
Galactic populations.  Thin- and thick-discs and halo stars have been
extracted from the entire {\sl APOGEE} dataset according to the rough
criteria after \citet[][$V_{\rm{tot}} \leq 50$\,km/s, $70 \leq V_{\rm{tot}}
\leq 180$\,km/s, and $V_{\rm{tot}} > 200$\,km/s, for the thin-disc,
thick-disc, and halo, respectively, where $V_{\rm{tot}} = (U^2 + V^2 +
W^2)^{0.5}$]{Ben2014}.  For the bulge stars, we adopted a very strict
criterion, taking into account only objects not more than 10\degr\, from
the Galactic center ($d \leq (l^2 + b^2)^{0.5}$).  The populations of thin-
and thick-discs and bulge permeate each other to some degree and overlap in
the diagram.  The positions corresponding to our SySts indicate that most of
them belong to the disc or bulge populations with a few halo candidates. 
Nevertheless, a more detailed analysis beyond the scope of this paper,
including, in particular, an exploration of the Toomre diagram, would be
needed to fully explore the kinematics of these objects.

%

An interesting by-product of this study is radial velocities for the giant
star and an additional component of the AG\,Peg system.  We derived the RG
radial velocity V$_{\rm Rad}$\,(A) = $-19.6 \pm 1.1$ at HJD\,2456904.94,
which corresponds very well with the radial velocity curve by
\citet{Fek2000a}.  According to their final ephemeris (Max($V$)=$JD\,2446812
+ 818.2 \times$\,E) the phase of this observation is $\phi = 0.336$, shortly
before the last photometric minimum preceding the Z\,And type outburst that
occurred in June\,2015 with the light maximum between July 2-nd and 14-th
\citep[see][]{Tom2016} which corresponds to $\phi \approx 0.71$.  The
second, blue-shifted component we observed has a radial velocity V$_{\rm
Rad}$\,(B) = $-37.3 \pm 2.2$\,km\,s${-1}$.  This could possibly indicate a
wind/an expanding circumstellar shell such as that discussed by
\citet{Qui2002} in the case of AR\,Pav, and perhaps can be connected in some
way with accretion on to the hot component leading to the outburst observed
roughly 300 days later.  Nevertheless, it should be noted that the cF
absorption system in AR\,Pav is in antiphase with the RG while the
observed radial velocity of the blue-shifted component in the case of our
observation of AG\,Peg is evidently not.  Accordingly, we cannot give a
confident interpretation without additional spectroscopic data collected
close in phase to this phenomenon.

\section*{Acknowledgements}

This study has been supported in part by the Polish National Science Center
(NCN) grant SONATA no.  DEC-2015/19/D/ST9/02974 and grant OPUS no. 
2017/27/B/ST9/01940.  Based on observations at Kitt Peak National
Observatory at NSF's NOIRLab (NOIRLab Prop.  ID 2014B-0240; PI: J. 
Miko{\l}ajewska), which is managed by the Association of Universities for
Research in Astronomy (AURA) under a cooperative agreement with the National
Science Foundation.  The authors are honored to be permitted to conduct
astronomical research on Iolkam Du'ag (Kitt Peak), a mountain with
particular significance to the Tohono O'odham.  This work has made use of
data from the European Space Agency (ESA) mission {\it Gaia}
(\url{https://www.cosmos.esa.int/gaia}), processed by the {\it Gaia} Data
Processing and Analysis Consortium (DPAC,
\url{https://www.cosmos.esa.int/web/gaia/dpac/consortium}).  Funding for the
DPAC has been provided by national institutions, in particular the
institutions participating in the {\it Gaia} Multilateral Agreement.

\section*{Data Availability}

The observed Phoenix spectral images utilized in this paper are 
available through the Gemini Science Archive, https://archive.gemini.edu/searchform.
The archive is searchable by object name. Reduced spectra are shown below.




\begin{thebibliography}{99}

  \bibitem[\protect\citeauthoryear{Ananth \& Leahy}{1993}]{AnLe1993} Ananth, A. G., \& Leahy, D. A., 1993,
     JApA, 14, 37

  \bibitem[\protect\citeauthoryear{Asplund et al.}{2009}]{Asp2009} Asplund, M., Grevesse, N., Sauval, A., Scott, P., 2009,
     ARA\&A, 47, 481

  \bibitem[\protect\citeauthoryear{Bailer-Jones et al.}{2018}]{Ba-J2018} Bailer-Jones, C. A. L., Rybizki, J., Fouesneau, M., et al., 2018, 
     AJ, 156, 58

  \bibitem[\protect\citeauthoryear{Belczy\'nski \& Miko{\l}ajewska}{1998}]{BeMi1998} Belczy\'nski, K., \&  Miko{\l}ajewska, J., 1998, 
     MNRAS, 296, 77

  \bibitem[\protect\citeauthoryear{Belczy\'nski et al.}{2000}]{Bel2000} Belczy\'nski, K., Miko{\l}ajewska, J., Munari, U., et al., 2000,
     A\&AS, 146, 407

  \bibitem[\protect\citeauthoryear{Bensby et al.}{2014}]{Ben2014} Bensby, T., Feltzing, S., \& Oey, M. S., 2014
     A\&A, 562, A71

  \bibitem[\protect\citeauthoryear{Bessell \& Wood}{1984}]{BeWo1984} Bessell, M. S., \& Wood, P. R., 1984,
     PASP, 96, 247

  \bibitem[\protect\citeauthoryear{Bessell \& Brett}{1988}]{BeBr1988} Bessell, M. S., \& Brett, J. M., 1988,
     PASP, 100, 1134

  \bibitem[\protect\citeauthoryear{Bessell et al.}{1998}]{Bes1998} Bessell, M. S., Castelli, F., \& Plez, B., 1998,
     A\&A, 333, 231.

  \bibitem[\protect\citeauthoryear{Boffin et al.}{2014}]{Bof2014} Boffin, H. M. J., Hillen, M., Berger, J. P., Jorissen, A., Blind, N., Le Bouquin, J. B., Miko{\l}ajewska, J., Lazareff, B., 2014,
     A\&A, 564, 1

  \bibitem[\protect\citeauthoryear{Brandt}{1998}]{Bra1998} Brandt, S., 1998,
     Data Analysis, Statistical and Computational Methods, Polish edn. Polish Scientific Publishers PWN, Warsaw

  \bibitem[\protect\citeauthoryear{Charbonnel \& Zahn}{2007}]{CaZ2007} Charbonnel, C., \& Zahn, J.-P., 2007,
     A\&A, 467, L15

  \bibitem[\protect\citeauthoryear{Chochol et al.}{1998}]{Cho1998} Chochol, D., Pribulla, T., \& Tamura, S., 1998,
     IBVS No. 4571, 1

  \bibitem[\protect\citeauthoryear{C\'uneo et al.}{2018}]{Cun2018} C\'uneo, V. A., Kenyon, S. J., G\'omez, M. N., Chochol, D., Shugarov, S. Y., \& Kolotilov, E. A., 2018,
     MNRAS, 479, 2728

  \bibitem[\protect\citeauthoryear{Dobrzycka et al.}{1993}]{Dob1993} Dobrzycka, D., Kenyon, S. J., Miko{\l}ajewska, J., 1993,
     AJ, 106, 284

  \bibitem[\protect\citeauthoryear{Eggleton}{1983}]{Eggl1983} Eggleton, P. P., 1983,
     ApJ, 268, 368

  \bibitem[\protect\citeauthoryear{Fekel et al.}{2000a}]{Fek2000a} Fekel, F. C., Joyce, R. R., Hinkle, K. H., \& Skrutskie, M. F., 2000a,
     AJ, 119, 1375

  \bibitem[\protect\citeauthoryear{Fekel et al.}{2000b}]{Fek2000b} Fekel, F. C., Hinkle, K. H., Joyce, R. R., \& Skrutskie, M. F., 2000b,
     AJ, 120, 3255

  \bibitem[\protect\citeauthoryear{Fekel et al.}{2001}]{Fek2001} Fekel, F. C., Hinkle, K. H., Joyce, R. R., \& Skrutskie, M. F., 2001,
     AJ, 121, 2219

  \bibitem[\protect\citeauthoryear{Fernie}{1985}]{Fer1985} Fernie, J. D., 1985,
     PASP, 97, 653

  \bibitem[\protect\citeauthoryear{Gaia Collaboration}{2018}]{Gai2018} Gaia Collaboration; Brown, A. G. A., Vallenari, A., Prusti, T., et al., 2018,
     A\&A, 616, 10 

  \bibitem[\protect\citeauthoryear{Gaia Collaboration}{2021}]{Gai2021} Gaia Collaboration; Brown, A. G. A., Vallenari, A., Prusti, T., et al. 2021,
     A\&A, 649, 1

  \bibitem[\protect\citeauthoryear{Gaia Collaboration}{2023}]{Gai2023} Gaia Collaboration; Vallenari, A., Brown, A. G. A., Prusti, T., et al. 2023,
     A\&A, 674, 1

  \bibitem[\protect\citeauthoryear{Ga{\l}an et al.}{2016}]{Gal2016} Ga{\l}an, C., Miko{\l}ajewska, J., Hinkle, K. H., Joyce, R. R., 2016,
     MNRAS, 455, 1282
 
  \bibitem[\protect\citeauthoryear{Ga{\l}an et al.}{2017}]{Gal2017} Ga{\l}an, C., Miko{\l}ajewska, J., Hinkle, K. H., Joyce, R. R., 2017,
     MNRAS, 466, 2194

  \bibitem[\protect\citeauthoryear{Glass \& Evans}{2003}]{GlEv2003}Glass, I. S., \& Evans, T. L., 2003,
     MNRAS, 343, 67

  \bibitem[\protect\citeauthoryear{Gonz{\'a}lez-Riestra et al.}{1990}]{Gon1990} Gonz{\'a}lez-Riestra, R., Cassatella, A., \& Fern{\'a}ndez-Castro, T., 1990,
     A\&A, 237, 385

  \bibitem[\protect\citeauthoryear{Goorvitch}{1994}]{Goo1994} Goorvitch, D., 1994,
     ApJS, 95, 535

  \bibitem[\protect\citeauthoryear{Gustafsson et al.}{2008}]{Gus2008} Gustafsson B., Edvardsson B., Eriksson K., J{\o}rgensen U.  G., Nordlund \AA, Plez B., 2008,
     A\&A, 486, 951

  \bibitem[\protect\citeauthoryear{Hachisu et al.}{1999}]{Hac1999} Hachisu, I., Kato, M., Nomoto, K., 1999,
     ApJ, 522, 487;

  \bibitem[\protect\citeauthoryear{Hidalgo et al.}{2018}]{Hid2018} Hidalgo S. L., Pietrinferni A., Cassisi S., et al., 2018
     ApJ, 856, 125

  \bibitem[\protect\citeauthoryear{Hinkle et al.}{1993}]{Hin1993} Hinkle K. H., Fekel F. C., Johnson D. S., \& Scharlach W. W. G., 1993, 
     AJ, 105, 1074

  \bibitem[\protect\citeauthoryear{Hinkle et al.}{1998}]{Hin1998} Hinkle K. H., Cuberly R. W., \& Gaughan N. A., et al., 1998, 
     Proc. SPIE, 3354, 810

  \bibitem[\protect\citeauthoryear{Hinkle et al.}{2009}]{Hin2009} Hinkle, K. H., Fekel, F. C., Joyce, R. R., 2009,
     AJ, 692, 1360

  \bibitem[\protect\citeauthoryear{Iijima et al.}{2019}]{Iij2019} Iijima, T., Naito, H., Narusawa, S., 2019,
     A\&A, 622, 45
 

  \bibitem[\protect\citeauthoryear{Ikeda \& Tamura}{2000}]{IkTa2000} Ikeda, Y., \& Tamura, S., 2000,
     PASJ, 52, 589

  \bibitem[\protect\citeauthoryear{Isogai et al.}{2010}]{Iso2010} Isogai, M., Seki, M., Ikeda, Y., Akitaya, H., \& Kawabata, K. S., 2010,
     AJ, 140, 235

  \bibitem[\protect\citeauthoryear{Jarrett et al.}{2011}]{Jar2011} Jarrett, T. H., Cohen, M., Masci, F. et al., 2011,
     ApJ, 735, 112.

  \bibitem[\protect\citeauthoryear{J\"onsson et al.}{2020}]{Jon2020} J\"onsson, H., Holtzman, J. A., Allende Prieto, C. et al, 2020,
     AJ, 160, 120

  \bibitem[\protect\citeauthoryear{Joyce}{1992}]{Joy1992} Joyce R., 1992, 
     in Howell S., ed., ASP Conf. Ser. Vol. 23, Astronomical CCD Observing and Reduction Techniques. Astron. Soc. Pac., San Francisco, p. 258

  \bibitem[\protect\citeauthoryear{J\"onsson et al.}{2020}]{Jon2020} J\"onsson, H., Holtzman, J. A., Allende Prieto, C., Cunha, K', et al., 2020,
     AJ, 160, 120

  \bibitem[\protect\citeauthoryear{Kato et al.}{2012}]{Kat2012} Kato, M., Miko{\l}ajewska, J., Hachisu, I., 2012,
     AJ, 750, 5

  \bibitem[\protect\citeauthoryear{Kato et al.}{2012b}]{Kat2012b} Kato, M., Miko{\l}ajewska, J., Hachisu, I., 2012b,
     Baltic Astronomy 21, 157

  \bibitem[\protect\citeauthoryear{Kenyon}{1986}]{Ken1986} Kenyon, S. J., 1986,
     The Symbiotic Stars (Cambridge: Cambridge Univ. Press)

  \bibitem[\protect\citeauthoryear{Kenyon}{1988}]{Ken1988} Kenyon, S. J., 1988,
     AJ, 96, 337

  \bibitem[\protect\citeauthoryear{Kenyon et al.}{1991}]{Ken1991} Kenyon, S. J., Oliversen, N. A., Miko{\l}ajewska, J., Miko{\l}ajewski, M., Stencel, R. E., Garcia, M. R., Anderson, C. M., 1991,
     AJ, 101, 63

  \bibitem[\protect\citeauthoryear{Kenyon et al.}{1993}]{Ken1993} Kenyon, S. J., Miko{\l}ajewska, J., Miko{\l}ajewski, M., Polidan, R, S., Slovak, M. H., 1993,
     AJ, 106, 1573

  \bibitem[\protect\citeauthoryear{Kenyon \& Garcia}{2016}]{KeGa2016} Kenyon, S. J., \& Garcia, M. R., 2016,
     AJ, 152, 1

  \bibitem[\protect\citeauthoryear{Kovetz \& Prialnik}{1997}]{KoPr1997} Kovetz A., \& Prialnik D., 1997,
     ApJ, 477, 356

  \bibitem[\protect\citeauthoryear{Kraft}{1958}]{Kra1958} Kraft, R. P., 1958, 
     ApJ, 127, 625

  \bibitem[\protect\citeauthoryear{Kucinskas et al.}{2005}]{Kuc2005}  Kucinskas, A., Hauschildt, P. H., Ludwig, H.-G., Brott, I., Vansevi\v{c}ius, V., Lindegren, L., Tanab\'e, T.,\& Allard, F., 2005,
     A\&A, 442, 281

  \bibitem[\protect\citeauthoryear{Kupka et al.}{1999}]{Kup1999} Kupka, F., Piskunov, N., Ryabchikova, T. A., Stempels, H. C., Weiss, W. W., 1999,
     A\&AS, 138, 119

  \bibitem[\protect\citeauthoryear{Kurucz}{1999}]{Kur1999} Kurucz, R. L., 1999,
     Available at: http://kurucz.harvard.edu

  \bibitem[\protect\citeauthoryear{Lagarde et al.}{2017}]{Lag2017} Lagarde, N., Robin, A. C., Reyl\'e, C., \& Nasello, G., 2017, 
     A\&A, 601, 27

  \bibitem[\protect\citeauthoryear{Lagarde et al.}{2019}]{Lag2019} Lagarde, N., Reyl\'e, C., A. C. Robin, A. C., et al., 2019,
     A\&A, 621, 24

  \bibitem[\protect\citeauthoryear{L\"u et al.}{2008}]{Lu2008} L\"u, G., Zhu, C., Han, Z., Wang, Z., 2008, 
     ApJ, 683, 990

  \bibitem[\protect\citeauthoryear{Liu et al.}{2018}]{Liu2018} Liu, D., Wang, B., Ge, H., Chen, X., Han, Z., 2018,
     MNRAS 473, 5352

  \bibitem[\protect\citeauthoryear{Majewski et al.}{2017}]{Maj2017} Majewski, S. R., Schiavon, R. P., Frinchaboy, P. M., et al., 2017, 
     AJ, 154, 94

  \bibitem[\protect\citeauthoryear{M{\'{e}}lendez \& Barbuy}{1999}]{MeBa1999} M{\'{e}}lendez, J., \& Barbuy, B., 1999,
     ApJS, 124, 52


  \bibitem[\protect\citeauthoryear{Miko{\l}ajewska \& Kenyon}{1992}]{MiKe1992} Miko{\l}ajewska, J., \& Kenyon, S. J., 1992,
     AJ, 103, 579

  \bibitem[\protect\citeauthoryear{Miko{\l}ajewska \& Kenyon}{1996}]{MiKe1996} Miko{\l}ajewska, J., \& Kenyon, S. J., 1996,
     AJ, 112, 1659

  \bibitem[\protect\citeauthoryear{Miko{\l}ajewska}{2003}]{Mik2003} Miko{\l}ajewska, J., 2003, in Corradi R. L. M., Mikolajewska J., Mahoney T. J., eds, ASP Conf. Ser. Vol. 303, Symboitic Stars Probing Steller Evolution. 
     Astron. Soc. Pac., San Francisco, p. 9

  \bibitem[\protect\citeauthoryear{Miko{\l}ajewska et al.}{2003b}]{Mik2003b} Miko{\l}ajewska, J., Ivison, R. J., \& Omont, A., 2003b, 
     in Corradi R. L. M., Mikolajewska J., Mahoney T. J., eds, ASP Conf. Ser. Vol. 303, Symboitic Stars Probing Steller Evolution. Astron. Soc. Pac., San Francisco, p. 478

  \bibitem[\protect\citeauthoryear{Miko{\l}ajewska et al.}{2006}]{Mik2006} Miko{\l}ajewska, J., Friedjung, M., \& Quiroga, C., 2006,
     A\&A, 460, 191

  \bibitem[\protect\citeauthoryear{Miko{\l}ajewska et al.}{2010}]{Mik2010} Miko{\l}ajewska, J., Balega, Y., Hofmann, K.-H., \& Weigelt, G., 2010,
     MNRAS, 403, 21

  \bibitem[\protect\citeauthoryear{Miko{\l}ajewska}{2012}]{Mik2012} Miko{\l}ajewska, J., 2012,
     Baltic Astronomy, 21, 5

  \bibitem[\protect\citeauthoryear{Miko{\l}ajewska}{2013}]{Mik2013} Miko{\l}ajewska, J., 2013,
     in Di Stefano, R., Orio, M., Moe, M., eds, IAU S281, Binary Paths to Type Ia Supernovae Explosions, Cambridge U. Press, 2013, p. 162

  \bibitem[\protect\citeauthoryear{Miko{\l}ajewska et al.}{2014}]{Mik2014} Miko{\l}ajewska, J., Ga{\l}an, C., Hinkle, K. H., Gromadzki, M., Schmidt, M. R., 2014,
     MNRAS, 440, 3016

  \bibitem[\protect\citeauthoryear{Miko{\l}ajewska \& Shara}{2017}]{MiSh2017} Miko{\l}ajewska, J., \& Shara, M. M., 2017,
     ApJ, 847, 99;

  \bibitem[\protect\citeauthoryear{Munari \& Jurdana-{\v{S}}epi{\'c}}{2002}]{MuJu2002} Munari, U., \& Jurdana-{\v{S}}epi{\'c}, R., 2002,
     A\&A, 386, 237

  \bibitem[\protect\citeauthoryear{Munari}{2019}]{Mun2019} Munari, U., 2019, Camb. Astrophys. Ser., 54, 77

  \bibitem[\protect\citeauthoryear{M{\"u}rset \& Schmid}{1999}]{Mue1999}  M{\"u}rset, U., \& Schmid, H. M., 1999,
     A\&AS, 137, 473

  \bibitem[\protect\citeauthoryear{M{\"u}rset et al.}{2000}]{Mue2000}  M{\"u}rset, U., Dumm, T., Isenegger, S., Nussbaumer, H., Schild, H., Schmid, H. M., \& Schmutz, W., 2000,
     A\&A, 353, 952

  \bibitem[\protect\citeauthoryear{Nussbaumer et al.}{1988}]{Nus1988} Nussbaumer, H., Schild, H., Schmid, H.\,M., \& Vogel M., 1988, 
     A\&A, 198, 179

  \bibitem[\protect\citeauthoryear{Paczy{\'n}ski}{1971}]{Pac1971} Paczy{\'n}ski, B., 1971,
     ARAA, 9, 183

  \bibitem[\protect\citeauthoryear{Pereira}{1995}]{Per1995} Pereira C.\,B., 1995,
     A\&A, 111, 471

  \bibitem[\protect\citeauthoryear{Phillips}{2007}]{Phi2007} Phillips, J. P., 2007,
     MNRAS, 376, 1120

  \bibitem[\protect\citeauthoryear{Poyner}{2012}]{Poy2012} Poyner, G., 2012,
     JBAA, 122, 356

  \bibitem[\protect\citeauthoryear{Pribulla et al.}{2003}]{Pri2003} Pribulla, T., Chochol, D., \& Parimucha, \v{S}., 2003,
     in Corradi, R.  L.  M., Mikolajewska, J., Mahoney, T.  J., eds, ASP Conf.  Ser.  Vol.  303, Symbiotic Stars Probing Stellar Evolution, Astron.  Soc.  Pac., San Francisco, p. 245,  

  \bibitem[\protect\citeauthoryear{Price et al.}{2010}]{Pri2010} Price, S. D., Smith, B. J., Kuchar, T. A., Mizuno, D. R., \& Kraemer, K. E., 2010,
     ApJSS, 190, 203

  \bibitem[\protect\citeauthoryear{Quiroga et al.}{2002}]{Qui2002} Quiroga, C., Miko{\l}ajewska, J., Brandi, E., Ferrer, O., \& Garc\'ia, L., 2002,
     A\&A, 387, 139

  \bibitem[\protect\citeauthoryear{Richichi et al.}{1999}]{Ric1999} Richichi, A., Fabbroni, L., Ragland, S.,\& Scholz, M., 1999,
     A\&A, 344, 511

  \bibitem[\protect\citeauthoryear{Schmid \& Schild}{1990}]{ScSc1990} Schmid H. M., Schild H., 1990,
     MNRAS, 246, 84

  \bibitem[\protect\citeauthoryear{Schild \& Schmid}{1997}]{ScSc1997} Schild, H., \& Schmid, H. M., 1997,
     A\&A, 324, 606

  \bibitem[\protect\citeauthoryear{Schmid \& Schild}{1997}]{ScSc1997-2} Schmid, H. M., \& Schild, H., 1997,
     A\&A, 327, 219

  \bibitem[\protect\citeauthoryear{Schmidt et al.}{2006}]{Sch2006} Schmidt, M. R., Za{\v c}s, L., Miko{\l}ajewska, J., \& Hinkle, K. H., 2006,
     A\&A, 446, 603

  \bibitem[\protect\citeauthoryear{Schlafly \& Finkbeiner}{2011}]{Sch2011} Schlafly, E. F.,\& Finkbeiner, D. P., 2011,
     ApJ, 737, 103

  \bibitem[\protect\citeauthoryear{Schlegel, Finkbeiner \& Davis}{1998}]{Sch1998} Schlegel, D. J., Finkbeiner, D. P.,\& Davis, M., 1998,
     ApJ, 500, 525

  \bibitem[\protect\citeauthoryear{Scott et al.}{2015}]{Sco2015} Scott, P., Asplund, M., Grevesse, N., Bergemann, M., Sauval, A. J., 2015,
     A\&A, 573, 26

  \bibitem[\protect\citeauthoryear{Shugarov et al.}{2012}]{Shu2012} Shugarov, S., Chochol, D., \& Kolotilov, E., 2012,
     Baltic Astronomy, 21, 150

  \bibitem[\protect\citeauthoryear{Smith \& Lambert}{1988}]{SmLa1988} Smith, V. V., \& Lambert, D., 1988,
     ApJ, 333, 219

  \bibitem[\protect\citeauthoryear{Sneden et al.}{2014}]{Sne2014} Sneden, C., Lucatello, S., Ram, R. S., Brooke, J. S. A., Bernath, P., 2014,
     ApJS, 214, 26

  \bibitem[\protect\citeauthoryear{Stanishev et al.}{2004}]{Sta2004} Stanishev, V., Zamanov, R., Tomov, N., Marziani, P., 2004,
     A\&A, 415, 609

  \bibitem[\protect\citeauthoryear{Tatarnikova et al.}{2018}]{Tat2018} Tatarnikova, A. A., Tatarnikov, A. M., Kolitilov, E. A., Shenavrin, V. I., Komissarova, G. V., 2018,
     Astronomy Letters, Vol. 44, No. 12, 803

  \bibitem[\protect\citeauthoryear{Tomov et al.}{2016}]{Tom2016} Tomov, T. V., Stoyanov, K. A., Zamanov, R. K., 2016,
     MNRAS, 462, 4435

  \bibitem[\protect\citeauthoryear{Van Belle et al.}{1999}]{VBe1999} Van Belle, G. T., Lane, B. F.,  Thompson, R. R., Boden, A. F., Colavita, M. M., Dumont, P. J., Mobley, D. W., Palmer, D., Shao, M., Vasisht, G. X., Wallace,
     AJ, 117, 521

  \bibitem[\protect\citeauthoryear{Vogel et al.}{1992}]{Vog1992} Vogel, M., Nussbaumer, H., \& Monier, R., 1992, 
     A\&A, 260, 156

  \bibitem[\protect\citeauthoryear{Wilson \& Vaccaro}{1997}]{WiVa1997} Wilson, R. E., \& Vaccaro, T. R., 1997,
     MNRAS, 291, 54.

  \bibitem[\protect\citeauthoryear{Yudin et al.}{2005a}]{Yud2005a} Yudin, B. F., Shenavrin, V. I., Kolotilov, E. A., Tatarnikova, A. A., \& Tatarnikov, A. M., 2005a,
     Astron. Rep. 49, 232

  \bibitem[\protect\citeauthoryear{Yudin et al.}{2005b}]{Yud2005b} Yudin, B. F., Kolotilov, E. A., Shenavrin, V. I., Tatarnikova, A. A., \& Tatarnikov, A. M., 2005b,
     A\&AT, 24, 447

\end{thebibliography}




\newpage
\clearpage

\appendix

\section{Notes on particular objects} \label{sc.npo}

\subsection{\sl EG\,And}

\citet{KeGa2016} have refined the orbital parameters of EG\,And using new
spectroscopic data spanning 20 years.  They estimated the mass of the giant
in the range $M_{\rm{g}} = 1.1$ -- $2.4$\,M$_{\sun}$.  From observations of
eclipses in the $UV$ \citet{Vog1992} derived $R_{\rm{g}} = 74 \pm
10$\,R$_{\sun}$.  \citet{WiVa1997} have interpreted the $UBV$ light curve
variations in terms of ellipsoidal effect.  By fitting light curves
\citet{KeGa2016} showed that the red giant is almost filling the tidal
surface, and estimated its radius $R_{\rm{g}} \sim 100$--$230$\,R$_{\sun}$,
much larger than that obtained by \citet{Vog1992}.  The 2MASS
\citep{Phi2007} $K$ magnitude is $2.58 \pm 0.26$.  Like other bright 2MASS
sources, this measurement has a relatively large uncertainty.  However,
\citet{Ken1988} and \citet{AnLe1993} obtained similar values, $2.59 \pm
0.02$ and $2.68 \pm 0.18$, respectively.  The relatively small reddening
toward EG\,And ($E$($B-V$)$< 0.07$) yields an intrinsic $K$ magnitude ($K_0
\sim 2.60$).  Adopting the Gaia's DR\,3 distance 0.594\,kpc we derive
$R_{\rm{g}} \sim 110$\,R$_{\sun}$ (Table\,\ref{T-eBMaRg}).  The inclination
of the orbit is not well known with literature values in the wide range
$45\degr$ -- $90\degr$.  Here we adopt the value $i = 70\degr$, which is
compatible with the longest eclipse durations and is an upper limit from the
\citet{WiVa1997} model.

\subsection{\sl AX\,Per}

From a comprehensive analysis of spectroscopic and photometric observations,
\citet{MiKe1992} concluded that AX\,Per contains a red giant filling its
Roche lobe surface and transferring the matter onto the accretion disc
around its companion.  They obtained the orbital solution for both
components.  \citet{Fek2000b} improved the solution for the red giant by
including new measurements of radial velocities and obtained a more accurate
semi-amplitude for the cool component $K_{\rm{g}} = 7.81 \pm
0.21$\,km\,s$^{-1}$ that reduced slightly the value of mass ratio to $q =
2.3 \pm 0.3$.  These results indicate the mass of the RG in the system
$M_{\rm{g}} \sim 1.0$\,M$_{\sun}$.  For the giant filling its tidal lobe
\citet{MiKe1992} obtained the inclination $i = 70\degr \pm 3\degr$.  Their
estimate of the giant's radius $R_{\rm{g}} \sim 0.8$\,AU $= 172$\,R$_{\sun}$
remains in agreement with the value obtained with the use of the Gaia
parallax $R = 132^{+22}_{-22}$\,R$_{\sun}$ (Table\,\ref{T-eBMaRg}).

\subsection{\sl T\,CrB}

\citet{BeMi1998} have analyzed ellipsoidal variations in optical light
curves of T\,CrB together with the mass function and information on
rotational velocity ($V_{\rm{rot}} \sin{i}$) as well as taking into account
evolutionary limits on the component masses to constrain the parameters of
the system.  They concluded that T\,CrB is a low mass binary system in which
the giant is the less massive component ($M_{\rm{g}} = 0.7 \pm
0.2$\,M$_{\sun}$).  With the mass ratio $q \approx 0.6 \pm 0.2$, the WD mass
is $M_{\rm{h}} = 1.2 \pm 0.2$\,M$_{\sun}$.

\citet{Sta2004} have analyzed the radial velocity curves of both components
including also the velocity curve of the hot component derived from the
wings of H${\alpha}$ emission, and they obtained the values: $M_{\rm{g}} =
1.12 \pm 0.23$\,M$_{\sun}$, $M_{\rm{h}} = 1.37 \pm 0.13$\,M$_{\sun}$, and
the mass ratio $q \approx 0.82 \pm 0.10$.  \citet{ BeMi1998} argued that the
value of $q \geq 0.8$ is unlikely because it would make the semidetached
binary unstable, so we adopt here a mass ratio of approximately $q \sim 0.7$
and a resultant red giant mass close to $M_{\rm{g}} \sim 0.9$\,M$_{\sun}$. 
Eclipses of the hot component are not observed, and \citet{Sta2004} and
\citet{BeMi1998} both find the limit of inclination $i $<$ 70\degr$. 
\citet{BeMi1998} in their model adopted $i = 60\pm5\degr$.

\subsection{\sl FG\,Ser}

The spectroscopic orbit of the cool component in FG\,Ser was calculated by
\citet{Fek2000b} who added the data of \citet{Mue2000} to their set of
radial velocities obtained at the Kitt Peak Observatory.  The revised
semi-amplitude was then $K_{\rm{g}} = 6.92 \pm 0.26$\,km\,s$^{-1}$.  In
these papers the mass of the red giant was estimated as $M_{\rm{g}} \sim
1.7$\,M$_{\sun}$.  Using angular diameters ($0.83 \pm 0.03$ and
$0.94\pm0.05$\,mas) derived from infrared interferometry by \citet{Bof2014}
and assuming the distance from the Gaia DR\,3 parallax
(Table\,\ref{T-eBMaRg}) we estimated the stellar radius $R =
140^{+15}_{-13}$\,R$_{\sun}$.  This value is in perfect agreement with that
estimated from the parallax and infrared photometry (Table\,\ref{T-eBMaRg})
and remains consistent within the cited uncertainty with $R =
105\pm15$\,R$_{\sun}$ estimated by \citet{Mue2000} and with the value $R
\sim 116$\,R$_{\sun}$ which we can obtain from the orbital period
(Table\,\ref{T-MaR}) and rotational velocity (Table\,\ref{T-js}). 
\citet{Bof2014} showed that the giant in this system is filling its Roche
lobe.  They were able to measure variations in angular diameter with the
orbital phase which they interpreted as confirmation of ellipsoidal
variations, and concluded that the inclination of the orbit should be close
to being nearly edge-on.  FG\,Ser is a known eclipsing system
\citep{Mue2000}.  We adopt here $i = 90\degr$.

\subsection{\sl V443\,Her}

The spectroscopic orbit of the cool component was first analyzed by
\citet{Dob1993}.  \citet{Fek2000b} improved the solution including
additional data from Kitt Peak Observatory.  The obtained semi-amplitude is
$K_{\rm{g}} = 2.52 \pm 0.21$\,km\,s$^{-1}$, and the estimated mass of the
red giant is $M_{\rm{g}} \sim 2.5$\,M$_{\sun}$.  The binary is not
eclipsing, which means that the inclination must be less than $\sim
60-70\degr$.  \citet{Dob1993} suggested $i \sim 30\degr$ which seems to be a
reasonable lower limit.  The value of the red giant radius derived with the
use of the Gaia DR\,3 parallax and intrinsic $K_{0}$ magnitude is $R = 166
\pm 27$\,R$_{\sun}$ (Table\,\ref{T-eBMaRg}).  Assuming synchronous rotation
and taking into account the orbital period (Table\,\ref{T-MaR}) and the
radial velocity (Table\,\ref{T-js}) would give $R \sim 120$\,R$_{\sun}$.

\subsection{\sl V1413\,Aql}

V1413\,Aql is a symbiotic system of Z\,And type with an outburst observed in
the 1980s and subsequent maxima in 1993 and 1995.  The system remained in an
active phase until May\,2017 \citep{Tat2018}.  There is no known
spectroscopic orbit for any of the components, and we have no certain
information about the parameters.  We used a typical mass for symbiotic
giants and the distances from Gaia DR\,2, to estimate the surface gravity.

\subsection{\sl BF\,Cyg}

A comprehensive study of the spectroscopic orbit of the cool component in
BF\,Cyg was made by \citet{Fek2001}.  They obtained the semi-amplitude
$K_{\rm{g}} = 6.72 \pm 0.24$\,km\,s$^{-1}$.  They also derived the
spectroscopic orbit for the hot component using velocities measured by
\citet{Gon1990} from emission lines in high-resolution $IUE$ spectra to
obtain the semiamplitude $K_{\rm{h}} = 24.3 \pm 4.9$\,km\,s$^{-1}$ giving a
mass ratio $q = 3.6 \pm 0.9$.  The resultant minimum mass of the giant is
$M_{\rm{g}} = 1.8 \pm 0.6$\,M$_{\sun}$.  \citet{Yud2005a, Yud2005b} based on
analysis of infrared $JK$, and $UBV$ photometry obtained a stellar radius of
about half the orbital semi-major axis, indicating a giant filling the Roche
surface.  They estimated the orbital inclination as low as $i \sim 70\degr$,
which would result in $M_{\rm{g}} \sim 2.2$\,M$_{\sun}$.  Thus the red giant
radius should be $R \leq 240$\,R$_{\sun}$.  The Gaia DR\,2 parallax provides
a slightly smaller radius $R_{\rm{g}} = 163^{+40}_{-37}$\,R$_{\sun}$
(Table\,\ref{T-eBMaRg}).

\subsection{\sl CH\,Cyg}

\citet{Hin2009} have refined orbital elements for the cool component of
CH\,Cyg.  They concluded that this is a binary system containing a mass
accreting WD -- the longest-period so far known ($P_{\rm{orb}} = 5689 \pm
47$\,d) in an S-type symbiotic system.  They obtained the semi-amplitude
$K_{\rm{g}} = 4.45 \pm 0.12$\,km\,s$^{-1}$.  Through confrontation of the
stellar parameters with the stellar evolution theory, the mass of the red
giant was estimated on $M_{\rm{g}} \sim 2$\,M$_{\sun}$.  Analysis of the
eclipse geometry yielded an estimated radius of the giant $R \sim
280$\,R$_{\sun}$ and the orbital inclination $i = 84\degr$.  \citet{Iij2019}
have recently derived masses based on assumption of a triple model system
hypothesis originally proposed by \citet{Hin1993}.  This model was then
verified and rejected by \citet{Hin2009}.  Unfortunately, the paper by
Iijima et al.  does not addressed any of the significant caveats about this
model raised in the second paper by Hinkle et al.  , as well as they do not
provide convincing radial velocity curve for the hot component which was
used to derive the masses.  Thus, we do not trust these estimates.\\ The
binary system was resolved for the first time (the components are separated
by $42 \pm 2$\,mas) by \citet{Mik2010}, giving a total binary mass of about
3.7 M$_{\sun}$, in good agreement with the mass resulting from the
spectroscopic orbit derived by \citet{Hin2009}.  The distance estimated by
\citet{Mik2010} $220^{+40}_{-28}$\,pc is consistent with the new value
$205^{+3}_{-4}$\,pc derived by Gaia DR\,3 (Table\,\ref{T-eBMaRg}) as well as
with the previous Hipparcos result $244^{+49}_{-35}$\,pc.  The 2MASS
magnitudes of CH\,Cyg, $J= 1.07 \pm 0.29$ and $K = -0.42 \pm 0.19$ are in
fairly good agreement with with $J = 1.14 \pm 0.09$, $K = -0.43 \pm 0.13$
\citep{AnLe1993} and $J \sim 492$\,Jy, $K \sim 953$\,Jy \citep{Pri2010},
which when converted with the use of calibrations by \citet{Jar2011} and
\citet{Bes1998} give equivalent magnitudes $J \sim 1.25$ and $K \sim -0.44$,
respectively.  Taking into account the small reddening toward CH\,Cyg
($E$($B-V$)$< 0.07$) we can estimate intrinsic $K$ magnitude, $K_0 \sim
-0.44$, and colour ($J-K$)$_0 \sim 1.56$.  Adopting the distance provided by
\citet{Mik2010} the radius of the giant is $R_{\rm{g}} = 188 \pm
44$\,R$_{\sun}$.  Finally, we adopt the stellar parameters of CH\,Cyg
($T_{\rm{eff}} = 3100 \pm 80$ and $\log{g} = 0.0$ -- Table\,\ref{T-est})
identical to those used for abundance analysis by \citet{Sch2006}.

\subsection{\sl QW\,Sge}

QW\,Sge is a very poorly studied symbiotic system with an orbital period
determined from optical photometry of $P_{\rm{orb}} = 390.5$\,d
\citep{MuJu2002}.  Because the orbital parameters, as well as the parameters
of the system components are not known, we can only roughly estimate the
mass and the radii of the giant.  Therefore like in the case of V1413\,Aql
we used a typical mass for symbiotic giants and the distances from Gaia
DR\,3 together with known $J$ and $K$ magnitudes, to estimate the surface
gravity.  We adopted the value $\log{g} = 0.5$
(Tables\,\ref{T-est}\,\&\,\ref{T-elogg}).

\subsection{\sl CI\,Cyg}

Orbital elements of the components of CI\,Cyg were first studied by
\citet{Ken1991}.  The spectroscopic orbit for the cool component was later
refined by \citet{Fek2000a}.  The orbital period ($P_{\rm{orb}} =
855.25$\,d) derived by \citet{Ken1991} from photometry remains in very good
agreement with the $P_{\rm{orb}} = 853.8 \pm 2.9$\,d determined
spectroscopically by \citet{Fek2000a}.  The semi-amplitudes of radial
velocity curves of both components are now well known ($K_{\rm{g}} = 6.7 \pm
0.3$\,km\,s$^{-1}$, $K_{\rm{h}} = 17.3 \pm 2.8$\,km\,s$^{-1}$) and they
imply the mass ratio $q = 2.6 \pm 0.6$ \citep{Mik2006}.  This results in the
mass of the giant $M_{\rm{g}} = 0.9$\,M$_{\sun}$ if the orbit is seen
edge-on ($i = 90\degr$) and could increase up to $M_{\rm{g}} \sim
1.27$\,M$_{\sun}$ if we adopt the inclination $i = 73\degr \pm 6\degr$ as
estimated by \citet{Ken1991}.  Using the Gaia DR\,3 parallax and intrinsic
$K_0$ magnitude we can derive the giant's radius $R =
197$\,$\pm37$\,R$_{\sun}$ (Table\,\ref{T-eBMaRg}).

\subsection{\sl PU\,Vul}

PU\,Vul is one of several known symbiotic novae in S-type SySt.  It is an
eclipsing binary with orbital period $P_{\rm{orb}} \sim 4900$\,d
\citep{Shu2012, Cun2018}.  \citet{Kat2012} used multicolour $UBVRI$ light
curves spanning 32 years \citep{Shu2012} which covers three eclipses (in
1980, 1994, and 2007) and the archival $IUE$ data to construct the composite
light-curve model, and estimated the mass ($M_{\rm{g}} = 0.8$\,M$_{\sun}$),
the radius ($R_{\rm{g}} = 335$\,R$_{\sun}$), and the distance to the system
as $d \sim 4.7$\,kpc.  These values remain in pretty good agreement with
$M_{\rm{g}} = 0.76$\,M$_{\sun}$ and $R_{\rm{g}} = 282$\,R$_{\sun}$ derived
previously by \citet{Cho1998}.  \citet{Kat2012} noted, however, that this
large radius may be that of a thick TiO atmosphere, transparent in the
$K$-band but opaque in the $V$-band, and the radius of the giant in the
infrared can be significantly smaller.  Using the bolometric luminosity --
period relation for LMC Mira variables by \citet{GlEv2003} they estimated
the giant luminosity (L = 3300\,L$_{\sun}$) and radius $R_{\rm{g}} = 187 \pm
12$\,R$_{\sun}$, which we adopt as more reasonable for our considerations. 
The value of the Gaia DR\,3 parallax together with the $K_0$ magnitude gives
the giant's radius $R_{\rm{g}} = 197^{+57}_{-52}$\,R$_{\sun}$
(Table\,\ref{T-eBMaRg}).

\subsection{\sl V1329\,Cyg}

V1329\,Cyg is another symbiotic nova in our sample.  The orbital inclination
derived from spectropolarimetric observations \citep{ScSc1997} is $i =
86\degr \pm 2\degr$.  From the analysis of the light minima, they obtained
an orbital period $P_{\rm{orb}} = 956.5$\,d.  The spectroscopic orbit
solution \citep{Fek2001} gives the red giant semi-amplitude, $K_{\rm{g}} =
7.85 \pm 0.26$\,km\,s$^{-1}$.  \citet{Pri2003} using the optical emission
line profiles published by \citet{IkTa2000} and the radial velocities of
$UV$ emission lines measured from archival $IUE$ spectra derived the
semi-amplitude for the hot component $K_{\rm{h}} = 22.3 \pm
2.0$\,km\,s$^{-1}$ which combined with that of the red giant yields the mass
ratio $q = 2.84 \pm 0.35$ and the masses for the components $M_{\rm{g}} =
2.02 \pm 0.51$\,M$_{\sun}$ and $M_{\rm{h}} = 0.71 \pm 0.14$\,M$_{\sun}$. 
Assuming synchronous rotation and $V_{\rm{rot}} \sin{i} = 10.3 \pm 0.5$
km\,s$^{-1}$ (Table\,\ref{T-js}) we obtain the radius $R = 195 \pm 10
$\,R$_{\sun}$.

\subsection{\sl AG\,Peg}

AG\,Peg is the third symbiotic nova in our sample.  The spectroscopic orbit
of the cool component was calculated by \citet{Fek2000a} who included all
previously existing radial velocity measurements in their solution.  They
obtained the eccentric orbit ($e = 0.110 \pm 0.039$, $\omega = 112\degr \pm
22\degr$) and an improved semiamplitude $K_{\rm{g}} = 5.44 \pm
0.20$\,km\,s$^{-1}$.  The orbital period derived from spectroscopic data
$P_{\rm{orb}} = 818.2$\,d is in very good agreement with the photometric
$P_{\rm{orb}} = 816.5$\,d \citep{Fer1985}.  Using the semi-amplitude of the
hot component $K_{\rm{h}} = 21.7 \pm 6.6$\,km\,s$^{-1}$ \citep{Ken1993}
yields a mass ratio of $q = 4 \pm 1.4$.  \citet{Mik2003} gives the masses of
the components $M_{\rm{g}} \geq 1.8$\,M$_{\sun}$ and $M_{\rm{h}} \geq
0.46$\,M$_{\sun}$, adopting an orbital inclination $i \leq 60\degr$.  Using
the angular radius ($0.50 \pm 0.02$\,mas) derived from infrared
interferometry by \citet{Bof2014} and the distance from the Gaia DR\,3
parallax we estimate the stellar radius $R = 137$\,$\pm11$\,R$_{\sun}$
(Table\,\ref{T-eBMaRg}).

\subsection{\sl Z\,And}

Z\,And is a prototypical symbiotic star \citep{Ken1986}.  The first
spectroscopic orbit of the cool component was derived by \citet{MiKe1996}. 
It was refined later by \citet{Fek2000b} who combined the radial velocity
data from \citet{MiKe1996} with their new measurements, and obtained the
semi-amplitude $K_{\rm{g}} = 6.73 \pm 0.22$\,km\,s$^{-1}$ and the orbital
period $P_{\rm{orb}} = 759.0$\,d.  The mass ratio is unknown but using the
known mass function $f(M_{\rm{g}}) = 0.024$ and the measured rotational
velocity we can estimate its lower limit as $q \gtrsim 1.5$--$1.2$.  For the
mass of the red giant, we adopt $M_{\rm{g}} = 2$\,M$_{\sun}$
\citep{Fek2000b}.  Adopting the Gaia DR\,3 parallax and intrinsic $K_0$
magnitude we obtain the giant’s radius $R = 136^{+18}_{-19}$\,R$_{\sun}$
(Table\,\ref{T-eBMaRg}).  \citet{MiKe1996} based on analysis of optical
observations suggested the range $i \sim 50\degr$ -- $70\degr$. 
Spectropolarimetric studies of the Raman scattered \ion{O}{vi} lines
indicate $i = 47\degr \pm 12\degr$ \citep{ScSc1997-2} and $i = 41\degr \pm
8\degr$ \citep{Iso2010} while the continuum polarization changes suggest $i
= 73\degr \pm 14\degr$ \citep{Iso2010}.  We choose $i = 60\degr$ as a good
compromise.

%
   \begin{figure}
   \centering
   \includegraphics[width=\hsize]{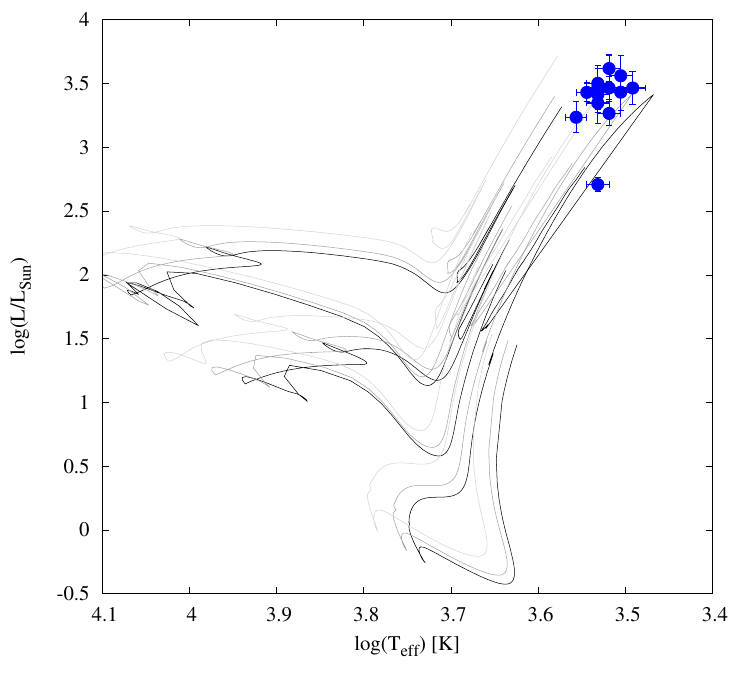}
      \caption{The H-R diagram with the positions of our targets (blue
points) compared to the {\sl{BaSTI}} evolutionary tracks \citep{Hid2018} for
stars of mass 1, 2 and 3 M$_{\sun}$, and three metallicities (Z = 0.0287 --
black, Z = 0.0172 (solar) -- dark-gray, and Z = 0.0077 -- light-gray).}
         \label{HR_A1}
   \end{figure}
%

\newpage

\section{Suplementary tables.} \label{AppT}

\begin{table*}
\caption[]{Estimation of bolometric magnitudes and stellar radii using
parallaxes from Gaia DR\,3 ({\sl{Top}}) \citep{Gai2023, Gai2021} and Gaia DR\,2
({\sl{Bottom}}) \citep{Gai2018}.}
\label{T-eBMaRg}
{\small{
\begin{tabular}{@{}l@{\hskip 3mm}l@{\hskip 3mm}c@{\hskip 3mm}c@{\hskip 3mm}c@{\hskip 3mm}c@{\hskip 3mm}c@{\hskip 3mm}c@{\hskip 3mm}l@{}}
\hline
            & $\varpi$         & {\sl{gofAL}}$^{a}$ & distance                  & $K_0$         & $M_{\rm K}$    & $BC_{\rm K}$  & $M_{\rm{Bol}}$ & $R_{\rm{rg}}$$^{b}$    \\
            & [mas]            &                    & [kpc]                     & mag           & mag            & mag           & mag            & [R$_{\sun}$]           \\
  \hline
 EG\,And    & 1.645$\pm$0.034  & 19.3               & 0.594$^{+0.014}_{-0.011}$ & $\sim 2.55$   & $\sim -6.3$    & $\sim 2.99$   & $\sim -3.3$    & $\sim 110$        \\
 AX\,Per    & 0.424$\pm$0.030  & 12.5               & 2.112$^{+0.155}_{-0.110}$ & 5.28$\pm$0.02 & -6.34$\pm$0.15 & 2.94$\pm$0.07 & -3.41$\pm$0.23 & 132$^{+22}_{-22}$ \\
 T\,CrB     & 1.092$\pm$0.028  & 25.6               & 0.887$^{+0.018}_{-0.029}$ & 4.79$\pm$0.02 & -4.95$\pm$0.08 & 2.94$\pm$0.06 & -2.01$\pm$0.14 & 65$^{+7}_{-8}$    \\
 FG\,Ser    & 0.625$\pm$0.046  & 1.06               & 1.472$^{+0.112}_{-0.090}$ & 4.12$\pm$0.04 & -6.72$\pm$0.19 & 2.94$\pm$0.10 & -3.78$\pm$0.29 & 147$^{+31}_{-30}$ \\
 V443\,Her  & 0.323$\pm$0.021  & 7.1                & 2.722$^{+0.150}_{-0.158}$ & 5.27$\pm$0.03 & -6.90$\pm$0.15 & 3.00$\pm$0.07 & -3.91$\pm$0.22 & 166$^{+27}_{-27}$ \\
 V1413\,Aql & 0.069$\pm$0.030  & 40.4               & 8.817$^{+2.162}_{-1.711}$ & 7.27$\pm$0.02 & -7.46$\pm$0.49 & 2.67$\pm$0.09 & -4.78$\pm$0.59 & 234$^{+115}_{-87}$\\
 BF\,Cyg    & 0.194$\pm$0.020  & 23.4               & 4.621$^{+0.321}_{-0.363}$ & 6.17$\pm$0.03 & -7.15$\pm$0.19 & 2.97$\pm$0.07 & -4.19$\pm$0.26 & 178$^{+33}_{-33}$ \\
 CH\,Cyg    & 4.882$\pm$0.087  & 13.0               & 0.205$^{+0.003}_{-0.004}$ & $\sim -0.44$  & $\sim -7.00$   & $\sim 3.24$   & $\sim -3.75$   & $\sim 145$        \\
 QW\,Sge    & 0.202$\pm$0.027  & 1.10               & 4.852$^{+0.576}_{-0.470}$ & 6.73$\pm$0.03 & -6.70$\pm$0.26 & 2.79$\pm$0.11 & -3.91$\pm$0.37 & 156$^{+42}_{-38}$ \\
 CI\,Cyg    & 0.476$\pm$0.026  & 2.8                & 1.944$^{+0.095}_{-0.096}$ & 4.24$\pm$0.04 & -7.20$\pm$0.15 & 2.91$\pm$0.12 & -4.29$\pm$0.26 & 197$^{+37}_{-37}$ \\
 PU\,Vul    & 0.191$\pm$0.040  & 38.6               & 4.234$^{+0.607}_{-0.549}$ & 6.01$\pm$0.02 & -7.12$\pm$0.32 & 2.98$\pm$0.08 & -4.15$\pm$0.40 & 197$^{+57}_{-52}$ \\
 V1329\,Cyg & 0.185$\pm$0.028  & 18.6               & 4.941$^{+0.751}_{-0.516}$ & 6.62$\pm$0.02 & -6.85$\pm$0.29 & 3.02$\pm$0.06 & -3.83$\pm$0.36 & 170$^{+44}_{-41}$ \\
 AG\,Peg    & 0.754$\pm$0.036  & 9.6                & 1.272$^{+0.051}_{-0.049}$ & 3.82$\pm$0.02 & -6.70$\pm$0.11 & 2.88$\pm$0.08 & -3.82$\pm$0.18 & 141$^{+19}_{-20}$ \\
 Z\,And     & 0.487$\pm$0.022  & 14.0               & 1.921$^{+0.083}_{-0.075}$ & 4.82$\pm$0.02 & -6.60$\pm$0.11 & 2.99$\pm$0.07 & -3.61$\pm$0.18 & 136$^{+18}_{-19}$ \\
  \hline
 EG\,And    & 1.486$\pm$0.039  &  18.8              & 0.660$^{+0.018}_{-0.016}$ & $\sim 2.55$   & $\sim -6.5$    & $\sim 2.99$   & $\sim -3.56$   & $\sim 120$        \\
 AX\,Per    & 0.298$\pm$0.057  &  17.3              & 2.924$^{+0.574}_{-0.423}$ & 5.28$\pm$0.02 & -7.05$\pm$0.38 & 2.94$\pm$0.07 & -4.11$\pm$0.46 & 182$^{+62}_{-54}$ \\
 T\,CrB     & 1.213$\pm$0.049  &  31.6              & 0.806$^{+0.034}_{-0.030}$ & 4.79$\pm$0.02 & -4.74$\pm$0.11 & 2.94$\pm$0.06 & -1.81$\pm$0.17 & 59$^{+8}_{-8}$    \\
 FG\,Ser    & 0.817$\pm$0.126  &  24.8              & 1.214$^{+0.242}_{-0.175}$ & 4.12$\pm$0.04 & -6.30$\pm$0.41 & 2.94$\pm$0.10 & -3.37$\pm$0.51 & 122$^{+48}_{-40}$ \\
 V443\,Her  & 0.472$\pm$0.041  &  36.1              & 2.000$^{+0.182}_{-0.155}$ & 5.27$\pm$0.03 & -6.24$\pm$0.21 & 3.00$\pm$0.07 & -3.24$\pm$0.28 & 122$^{+24}_{-24}$ \\
 V1413\,Aql & 0.140$\pm$0.038  &  14.8              & 5.557$^{+1.432}_{-0.984}$ & 7.27$\pm$0.02 & -6.45$\pm$0.48 & 2.67$\pm$0.09 & -3.78$\pm$0.57 & 147$^{+70}_{-54}$ \\
 BF\,Cyg    & 0.207$\pm$0.026  &   6.7              & 4.236$^{+0.534}_{-0.429}$ & 6.17$\pm$0.03 & -6.96$\pm$0.27 & 2.97$\pm$0.07 & -4.00$\pm$0.34 & 163$^{+40}_{-37}$ \\
 CH\,Cyg    & 5.464$\pm$0.217  & 154.8              & 0.183$^{+0.007}_{-0.007}$ & $\sim -0.44$  & $\sim -6.75$   & $\sim 3.24$   & $\sim -3.51$   & $\sim 130$        \\
 QW\,Sge    & 0.212$\pm$0.060  &  37.7              & 3.972$^{+1.222}_{-0.792}$ & 6.73$\pm$0.03 & -6.27$\pm$0.56 & 2.79$\pm$0.11 & -3.48$\pm$0.67 & 128$^{+82}_{-55}$ \\
 CI\,Cyg    & 0.560$\pm$0.050  &  26.0              & 1.716$^{+0.168}_{-0.141}$ & 4.24$\pm$0.04 & -6.93$\pm$0.23 & 2.91$\pm$0.12 & -4.02$\pm$0.35 & 174$^{+44}_{-41}$ \\
 PU\,Vul    & 0.520$\pm$0.088  &  66.4              & 1.851$^{+0.393}_{-0.279}$ & 6.01$\pm$0.02 & -5.33$\pm$0.41 & 2.98$\pm$0.08 & -2.35$\pm$0.49 & 86$^{+32}_{-27}$  \\
 V1329\,Cyg & 0.256$\pm$0.055  &  36.8              & 3.374$^{+0.768}_{-0.542}$ & 6.62$\pm$0.02 & -6.02$\pm$0.43 & 3.02$\pm$0.06 & -3.00$\pm$0.50 & 116$^{+44}_{-37}$ \\
 AG\,Peg    & 0.380$\pm$0.082  &   6.6              & 2.295$^{+0.506}_{-0.363}$ & 3.82$\pm$0.02 & -7.98$\pm$0.42 & 2.88$\pm$0.08 & -5.10$\pm$0.50 & 255$^{+98}_{-81}$ \\
 Z\,And     & 0.512$\pm$0.030  &  17.5              & 1.844$^{+0.110}_{-0.098}$ & 4.82$\pm$0.02 & -6.51$\pm$0.14 & 2.99$\pm$0.07 & -3.52$\pm$0.21 & 131$^{+20}_{-20}$ \\
  \hline
\end{tabular}
\begin{list}{}{}
\item[{\sl {Notes.} $^{a}${\sl{gofAL}} -- the goodness-of-fit statistic parameter describing the fit quality to the data of the Gaia measurements.
The better fit -- the smaller its value.}]
\item[$^b$]The radii are calculated using the values of $T_{\rm{eff}}$ as listed in the penultimate column of Table\,\ref{T-est}.
\end{list}
}}
\end{table*}

\begin{table}
 \centering
  \caption{Results of estimations on surface gravities {\sl {(II)}}, and its
limits {\sl {(I)}} and {\sl {(III)}} as described in the text
(Section\,\ref{sc.pg}).}
\label{T-elogg}
  \begin{tabular}{|@{}l@{\hskip 2mm}|l@{\hskip 2mm}|l@{\hskip 2mm}|l@{\hskip 2mm}|l|@{}}
  \hline
  Object   & \multicolumn{4}{c|}{$\log{g}$} \\
           & (\sl{I})& (\sl{II})              & (\sl{III})      & adopted \\
  \hline
EG\,And    & $-0.1<$ & ~$0.4$\,--\,$0.7$      & --              & 0.5     \\
AX\,Per    & $-0.1<$ & ~$0.05$\,--\,$0.35$    & $<0.3\pm0.2$    & 0.0     \\
T\,CrB     & ~$0.4<$ & $\sim0.7$              & $<0.3\pm0.2$    & 0.5     \\
FG\,Ser    & $-0.1<$ & ~$0.05$\,--\,$0.6$     & $<0.3\pm0.4$    & 0.5     \\
V443\,Her  & $-0.1<$ & ~$0.3$\,--\,$0.55$     & $<0.1\pm0.3$    & 0.5     \\
V1413\,Aql & --      & $-0.2$\,--\,$0.8$      & $<1.2\pm0.3$    & 0.5     \\
BF\,Cyg    & $-0.2<$ & $\geq 0.0$             & $<0.2\pm0.3$    & 0.0     \\
CH\,Cyg    & $-1.1<$ & \,$0.0$\,--\,$0.4$     & --              & 0.0     \\
QW\,Sge    & --      & $-0.15$\,--\,$0.6$     & $<0.8\pm0.4$    & 0.5     \\
CI\,Cyg    & $-0.2<$ & $-0.35$\,--\,$0.15$    & $<0.4\pm0.4$    & 0.0     \\
PU\,Vul    & --      & $\sim-0.2$             & $<0.2\pm0.3$    & 0.0     \\
V1329\,Cyg & $-0.2<$ & \,$0.0$\,--\,$0.3$     & $<$~$0.0\pm0.2$ & 0.0     \\
AG\,Peg    & $-0.3<$ & $\geq 0.35$\,--\,$0.5$ & $<0.5\pm0.3$    & 0.5     \\
Z\,And     & $-0.2<$ & ~$0.4$\,--\,$0.6$      & $<0.1\pm0.3$    & 0.5     \\
  \hline
\end{tabular}
\end{table}

\begin{table}
 \centering
  \caption{Sensitivity of abundances to uncertainties in the stellar
parameters.}
\label{TA-AbuSens}  
  \begin{tabular}{@{}lrrr@{}}

  \hline
  \hline
  $\Delta X$ & $\Delta T_{\rm{eff}} = +100$\,K & $\Delta \log{g} = +0.5$ & $\Delta  \xi_{\rm{t}} = +0.25$ \\
  \hline
  C          & $+0.04$ & $+0.21$ & $-0.03$ \\
  N          & $+0.03$ & $+0.00$ & $-0.05$ \\
  O          & $+0.12$ & $+0.07$ & $-0.05$ \\
  Sc         & $+0.13$ & $+0.11$ & $-0.27$ \\
  Ti         & $+0.09$ & $+0.12$ & $-0.23$ \\
  Fe         & $-0.04$ & $+0.14$ & $-0.07$ \\
  Ni         & $-0.06$ & $+0.17$ & $-0.12$ \\
  \hline
  \hline
\end{tabular}   
\end{table}

\newpage
\clearpage

\section{Figures -- spectra of 15 symbiotic giants observed in $K$-,
and/or $K_{\rm r}$-, and/or $H$-band regions, compared with synthetic fits.} \label{AppS}

\newpage

%
   \begin{figure}
   \centering
   \includegraphics[width=\hsize]{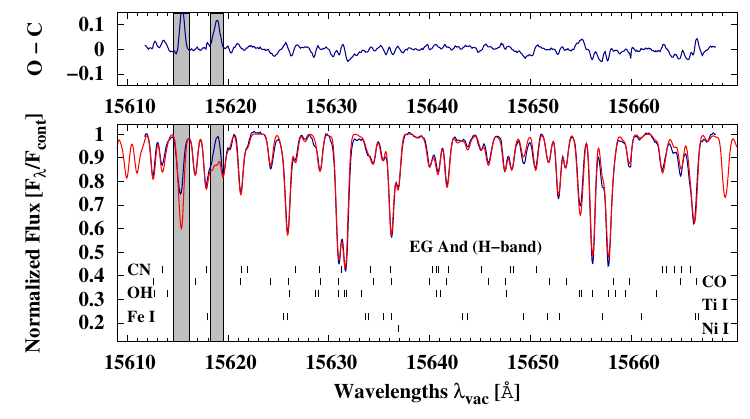}
      \caption{The $H$ band spectrum of EG\,And (blue line) and a synthetic
              spectrum (red line) calculated using the final abundances
              (Table\,\ref{T-finAbu}). The grey-shaded areas were
              excluded from calculations by a suitable mask.}
         \label{FC1}
   \end{figure}
%

%
   \begin{figure}
   \centering
   \includegraphics[width=\hsize]{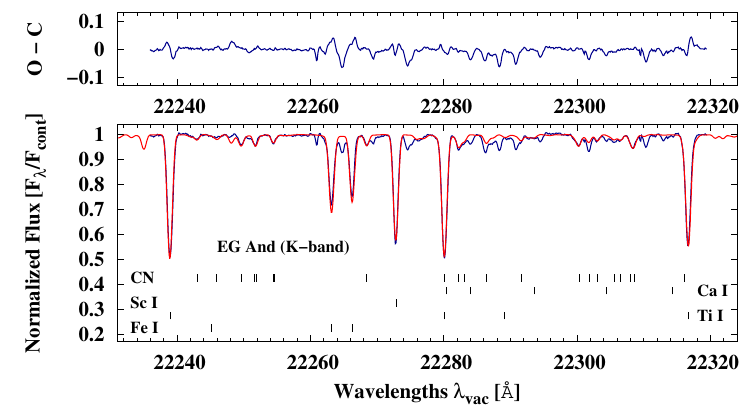}
      \caption{The $K$ band spectrum of EG\,And (blue line) and a synthetic
              spectrum (red line) calculated using the final abundances
              (Table\,\ref{T-finAbu}).}
         \label{FC2}
   \end{figure}
%

%
   \begin{figure}
   \centering
   \includegraphics[width=\hsize]{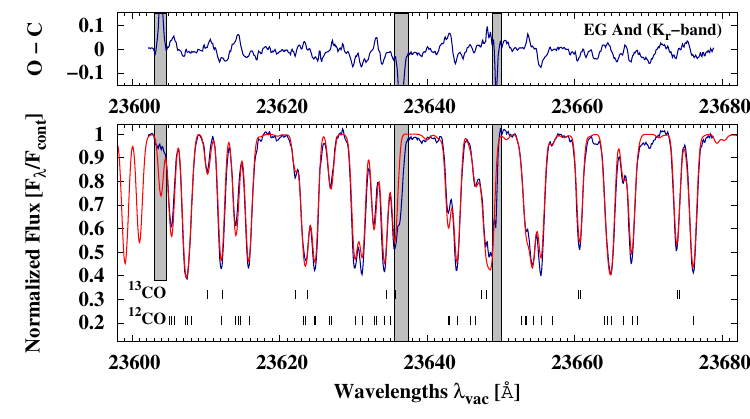}
      \caption{The $K_{\rm r}$ band spectrum of EG\,And (blue line) and a
              synthetic spectrum (red line) calculated using the final
              abundances (Table\,\ref{T-finAbu}). The grey-shaded areas
              were excluded from calculations by a suitable
              mask.}
         \label{FC3}
   \end{figure}
%

%
   \begin{figure}
   \centering
   \includegraphics[width=\hsize]{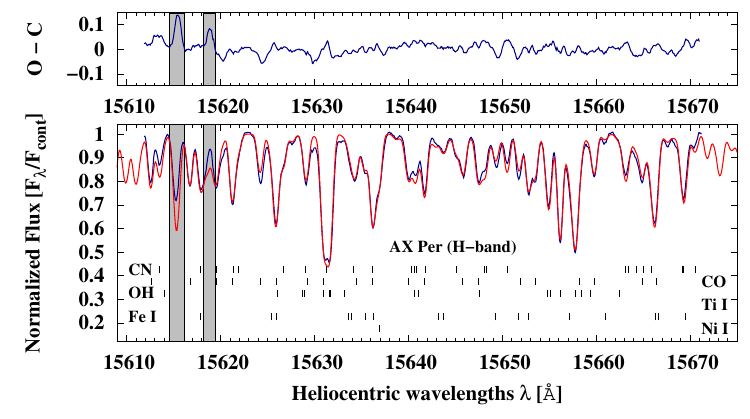}
      \caption{The $H$ band spectrum of AX\,Per (blue line) and a synthetic
              spectrum (red line) calculated using the final abundances
              (Table\,\ref{T-finAbu}).  The grey-shaded areas were
              excluded from calculations by a suitable mask.}
         \label{FC4}
   \end{figure}
%

%
   \begin{figure}
   \centering
   \includegraphics[width=\hsize]{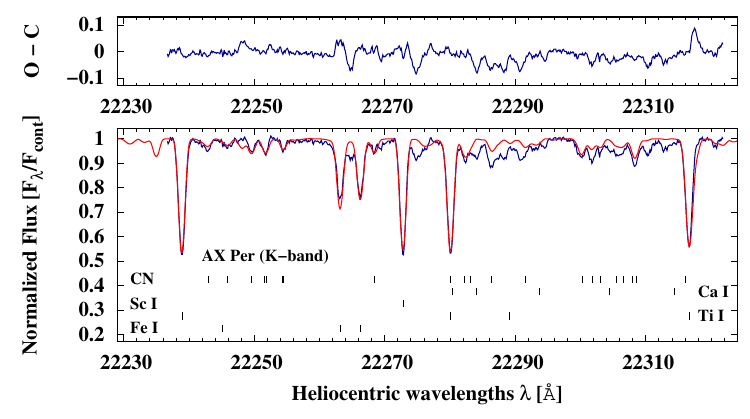}
      \caption{The $K$ band spectrum of AX\,Per (blue line) and a synthetic
              spectrum (red line) calculated using the final abundances
              (Table\,\ref{T-finAbu}).}
         \label{FC5}
   \end{figure}
%

%
   \begin{figure}
   \centering
   \includegraphics[width=\hsize]{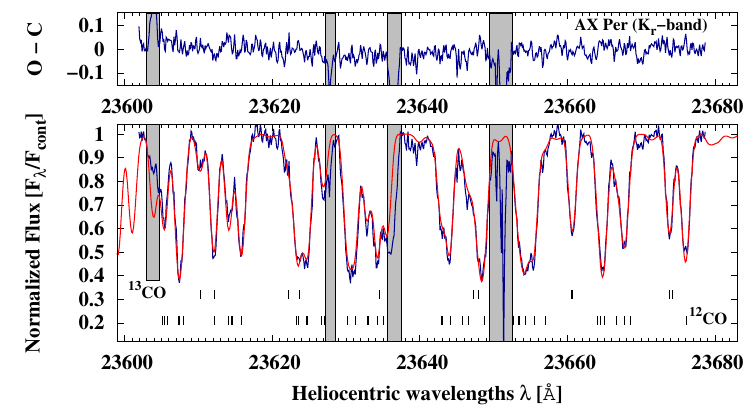}
      \caption{The $K_{\rm r}$ band spectrum of AX\,Per (blue line) and a
              synthetic spectrum (red line) calculated using the final
              abundances (Table\,\ref{T-finAbu}). The grey-shaded areas
              were excluded from calculations by a suitable
              mask.}
         \label{FC6}
   \end{figure}
%

%
   \begin{figure}
   \centering
   \includegraphics[width=\hsize]{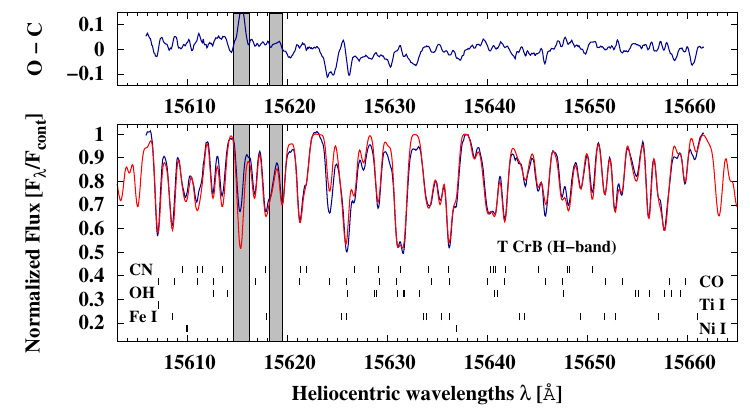}
      \caption{The $H$ band spectrum of T\,CrB (blue line) and a synthetic
              spectrum (red line) calculated using the final abundances
              (Table\,\ref{T-finAbu}). The grey-shaded areas were
              excluded from calculations by a suitable mask.}
         \label{FC7}
   \end{figure}
%

%
   \begin{figure}
   \centering
   \includegraphics[width=\hsize]{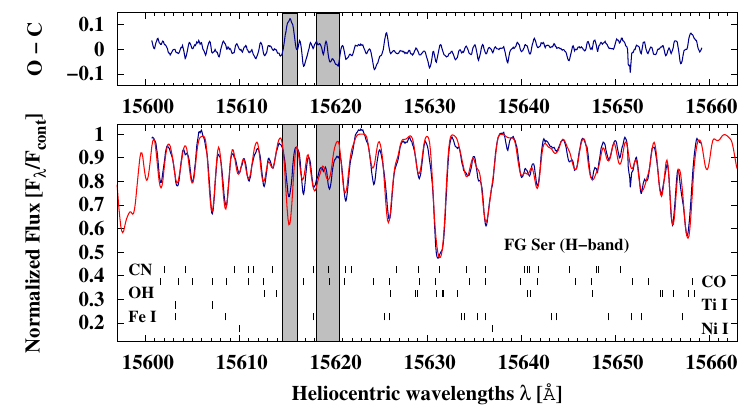}
      \caption{The $H$ band spectrum of FG\,Ser (blue line) and a synthetic
              spectrum (red line) calculated using the final abundances
              (Table\,\ref{T-finAbu}).  The grey-shaded areas were
              excluded from calculations by a suitable mask.}
         \label{FC8}
   \end{figure}
%

%
   \begin{figure}
   \centering
   \includegraphics[width=\hsize]{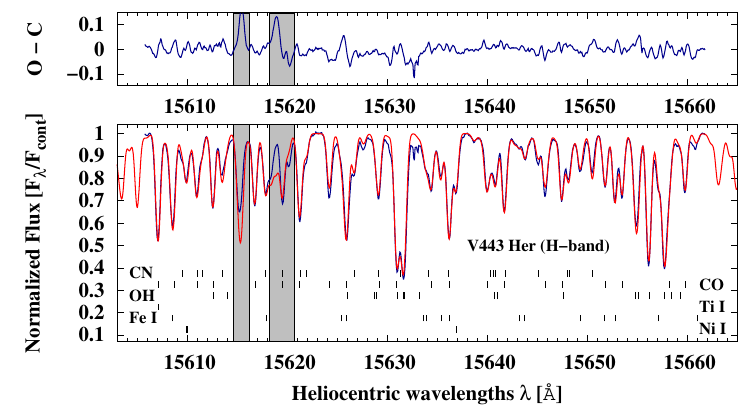}
      \caption{The $H$ band spectrum of V443\,Her (blue line) and a synthetic
              spectrum (red line) calculated using the final abundances
              (Table\,\ref{T-finAbu}).  The grey-shaded areas were
              excluded from calculations by a suitable mask.}
         \label{FC9}
   \end{figure}
%

%
   \begin{figure}
   \centering
   \includegraphics[width=\hsize]{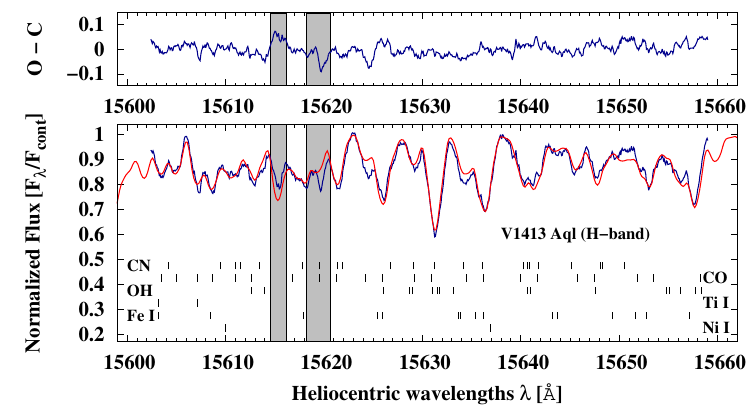}
      \caption{The $H$ band spectrum of V1413\,Aql (blue line) and a synthetic
              spectrum (red line) calculated using the final abundances
              (Table\,\ref{T-finAbu}).  The grey-shaded areas were
              excluded from calculations by a suitable mask.}
         \label{FC10}
   \end{figure}
%

%
   \begin{figure}
   \centering
   \includegraphics[width=\hsize]{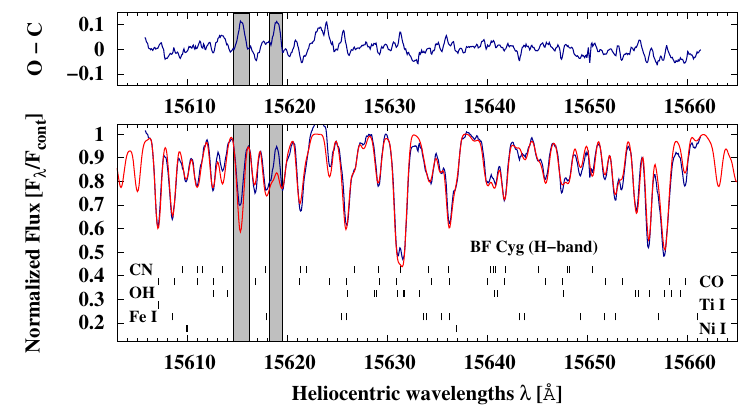}
      \caption{The $H$ band spectrum of BF\,Cyg (blue line) and a synthetic
              spectrum (red line) calculated using the final abundances
              (Table\,\ref{T-finAbu}). The grey-shaded areas were
              excluded from calculations by a suitable mask.}
         \label{FC11}
   \end{figure}
%

%
   \begin{figure}
   \centering
   \includegraphics[width=\hsize]{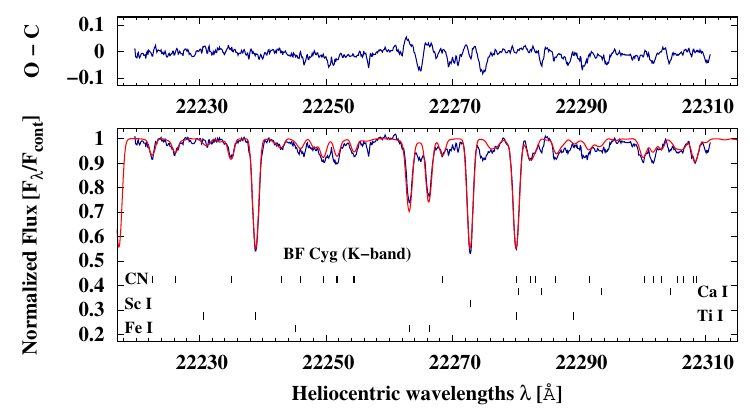}
      \caption{The $K$ band spectrum of BF\,Cyg (blue line) and a synthetic
              spectrum (red line) calculated using the final abundances
              (Table\,\ref{T-finAbu}).}
         \label{FC12}
   \end{figure}
%

%
   \begin{figure}
   \centering
   \includegraphics[width=\hsize]{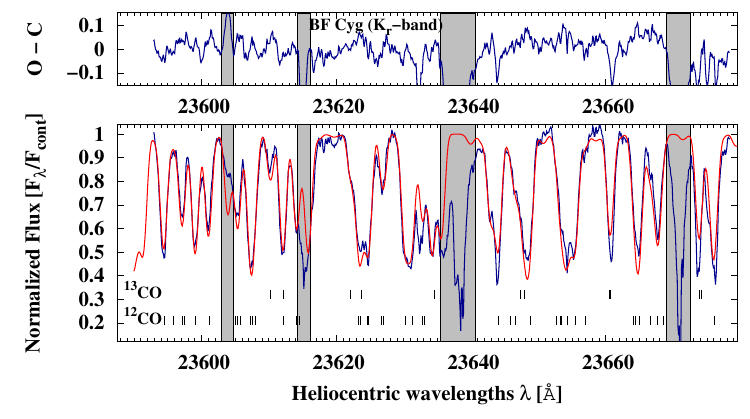}
      \caption{The $K_{\rm r}$ band spectrum of BF\,Cyg (blue line) and a
              synthetic spectrum (red line) calculated using the final
              abundances (Table\,\ref{T-finAbu}). The grey-shaded areas
              were excluded from calculations by a suitable
              mask.}
         \label{FC13}
   \end{figure}
%

%
   \begin{figure}
   \centering
   \includegraphics[width=\hsize]{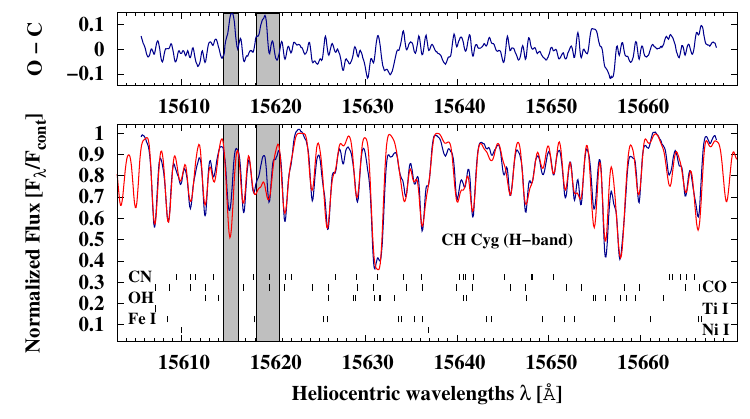}
      \caption{The $H$ band spectrum of CH\,Cyg (blue line) and a synthetic
              spectrum (red line) calculated using the final abundances
              (Table\,\ref{T-finAbu}). The grey-shaded areas
              were excluded from calculations by a suitable
              mask.}
         \label{FC14}
   \end{figure}
%

%
   \begin{figure}
   \centering
   \includegraphics[width=\hsize]{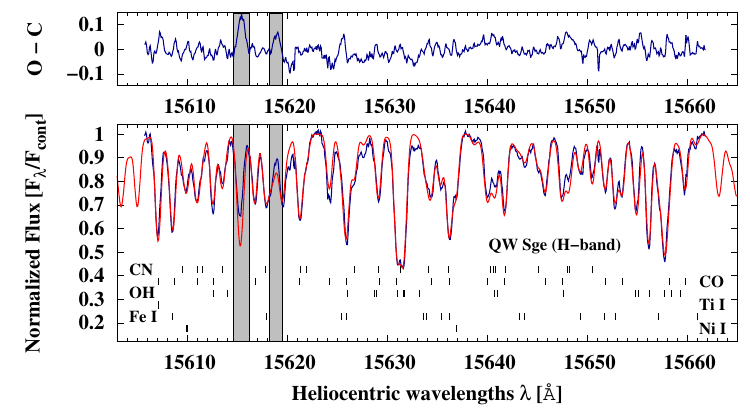}
      \caption{The $H$ band spectrum of QW\,Sge (blue line) and a synthetic
              spectrum (red line) calculated using the final abundances
              (Table\,\ref{T-finAbu}). The grey-shaded areas
              were excluded from calculations by a suitable
              mask.}
         \label{FC15}
   \end{figure}
%

%
   \begin{figure}
   \centering
   \includegraphics[width=\hsize]{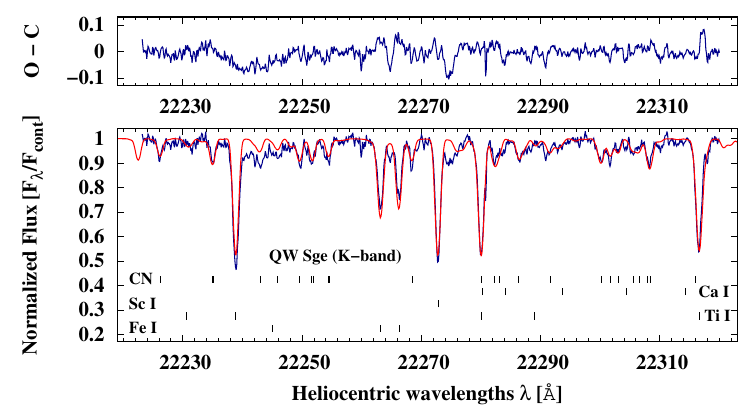}
      \caption{The $K$ band spectrum of QW\,Sge (blue line) and a synthetic
              spectrum (red line) calculated using the final abundances
              (Table\,\ref{T-finAbu}).}
         \label{FC16}
   \end{figure}
%

%
   \begin{figure}
   \centering
   \includegraphics[width=\hsize]{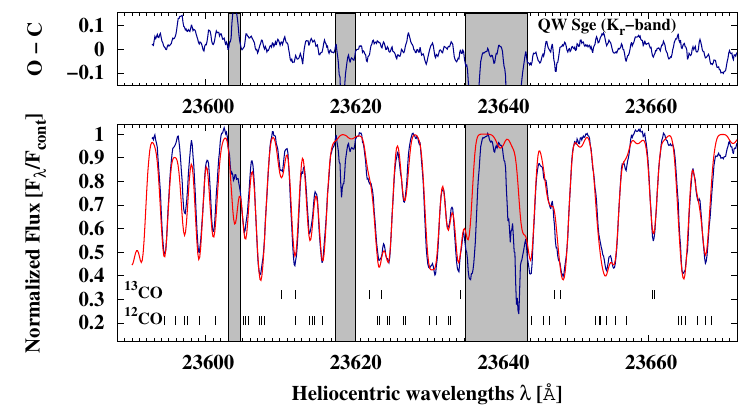}
      \caption{The $K_{\rm r}$ band spectrum of QW\,Sge (blue line) and a
              synthetic spectrum (red line) calculated using the final abundances
              (Table\,\ref{T-finAbu}). The grey-shaded areas
              were excluded from calculations by a suitable
              mask.}
         \label{FC17}
   \end{figure}
%

%
   \begin{figure}
   \centering
   \includegraphics[width=\hsize]{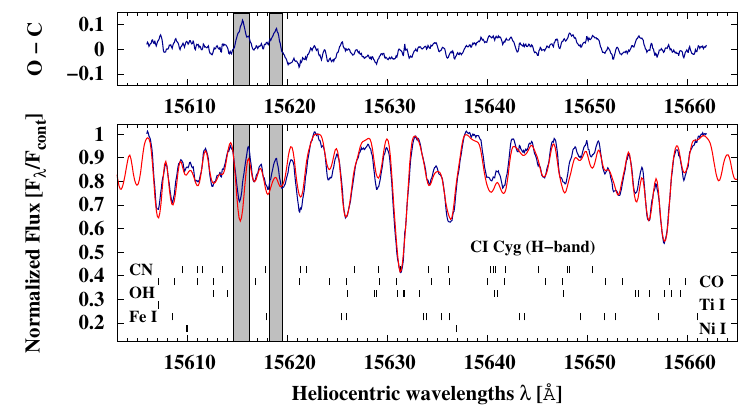}
      \caption{The $H$ band spectrum of CI\,Cyg (blue line) and a synthetic
              spectrum (red line) calculated using the final abundances
              (Table\,\ref{T-finAbu}). The grey-shaded areas were
              excluded from calculations by a suitable mask.}
         \label{FC18}
   \end{figure}
%

%
   \begin{figure}
   \centering
   \includegraphics[width=\hsize]{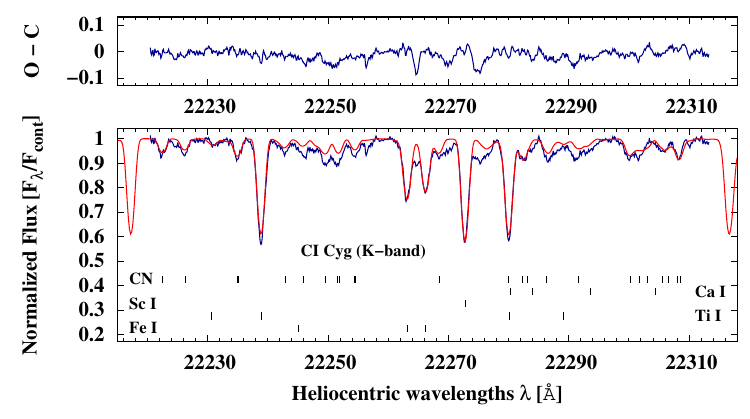}
      \caption{The $K$ band spectrum of CI\,Cyg (blue line) and a synthetic
              spectrum (red line) calculated using the final abundances
              (Table\,\ref{T-finAbu}).}
         \label{FC19}
   \end{figure}
%

%
   \begin{figure}
   \centering
   \includegraphics[width=\hsize]{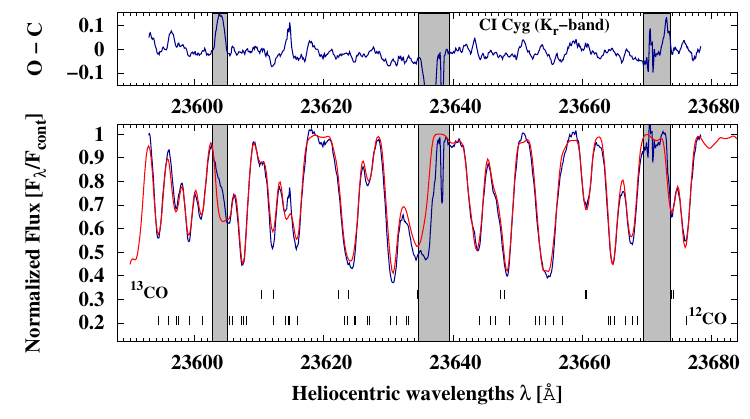}
      \caption{The $K_{\rm r}$ band spectrum of CI\,Cyg (blue line) and a
              synthetic spectrum (red line) calculated using the final
              abundances (Table\,\ref{T-finAbu}). The grey-shaded areas
              were excluded from calculations by a suitable
              mask.}
         \label{FC20}
   \end{figure}
%

%
   \begin{figure}
   \centering
   \includegraphics[width=\hsize]{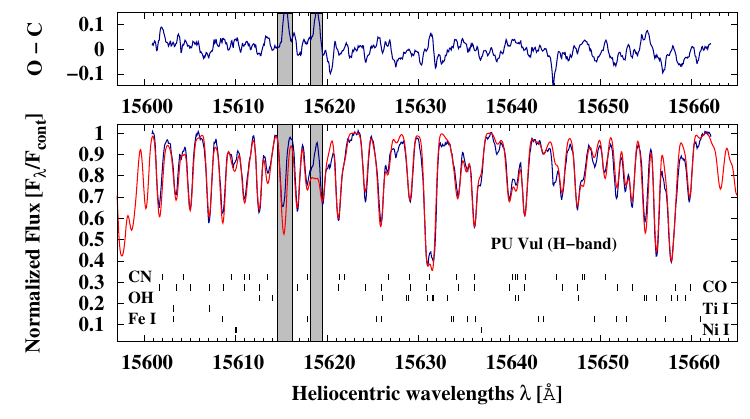}
      \caption{The $H$ band spectrum of PU\,Vul (blue line) and a synthetic
              spectrum (red line) calculated using the final abundances
              (Table\,\ref{T-finAbu}). The grey-shaded areas
              were excluded from calculations by a suitable
              mask.}
         \label{FC21}
   \end{figure}
%

%
   \begin{figure}
   \centering
   \includegraphics[width=\hsize]{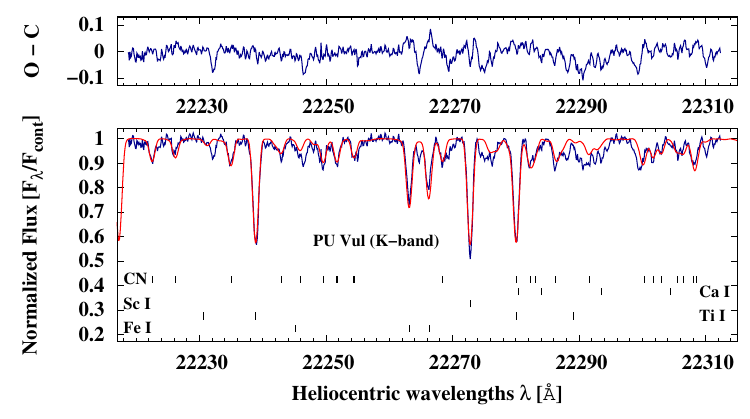}
      \caption{The $K$ band spectrum of PU\,Vul (blue line) and a synthetic
              spectrum (red line) calculated using the final abundances
              (Table\,\ref{T-finAbu}).}
         \label{FC22}
   \end{figure}
%

%
   \begin{figure}
   \centering
   \includegraphics[width=\hsize]{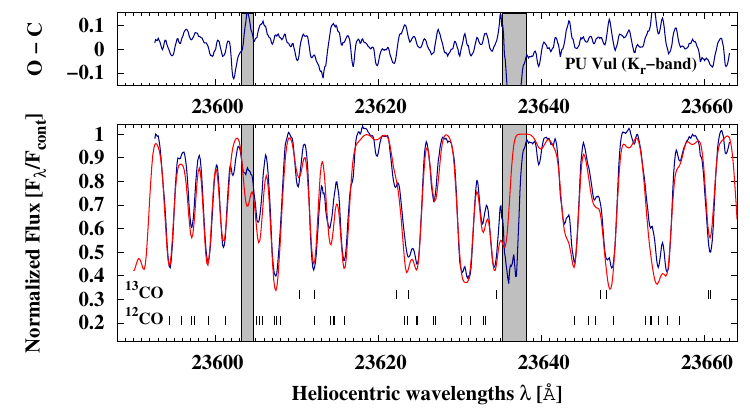}
      \caption{The $K_{\rm r}$ band spectrum of PU\,Vul (blue line) and a
              synthetic spectrum (red line) calculated using the final
              abundances (Table\,\ref{T-finAbu}). The grey-shaded areas
              were excluded from calculations by a suitable
              mask.}
         \label{FC23}
   \end{figure}
%

%
   \begin{figure}
   \centering
   \includegraphics[width=\hsize]{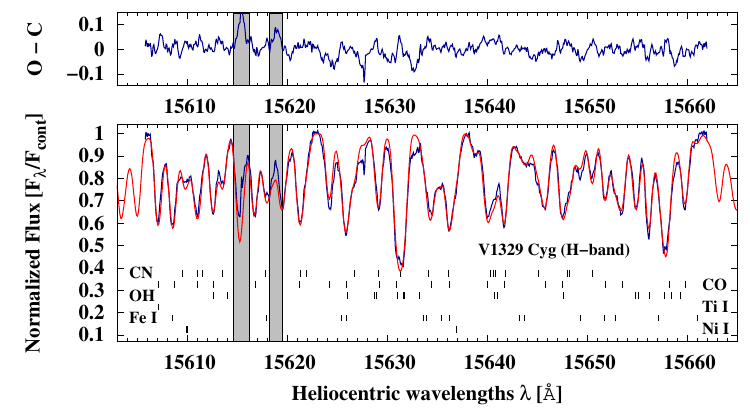}
      \caption{The $H$ band spectrum of V1329\,Cyg (blue line) and a synthetic
              spectrum (red line) calculated using the final abundances
              (Table\,\ref{T-finAbu}). The grey-shaded areas
              were excluded from calculations by a suitable
              mask.}
         \label{FC24}
   \end{figure}
%

%
   \begin{figure}
   \centering
   \includegraphics[width=\hsize]{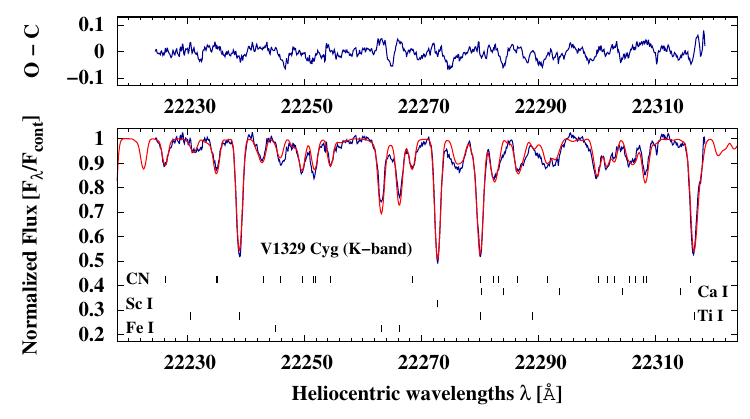}
      \caption{The $K$ band spectrum of V1329\,Cyg (blue line) and a synthetic
              spectrum (red line) calculated using the final abundances
              (Table\,\ref{T-finAbu}).}
         \label{FC25}
   \end{figure}
%

%
   \begin{figure}
   \centering
   \includegraphics[width=\hsize]{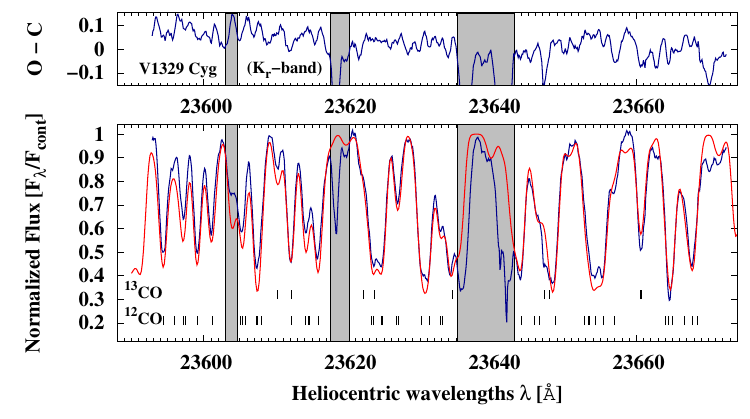}
      \caption{The $K_{\rm r}$ band spectrum of V1329\,Cyg (blue line) and a
              synthetic spectrum (red line) calculated using the final
              abundances (Table\,\ref{T-finAbu}). The grey-shaded areas
              were excluded from calculations by a suitable
              mask.}
         \label{FC26}
   \end{figure}
%

%
   \begin{figure}
   \centering
   \includegraphics[width=\hsize]{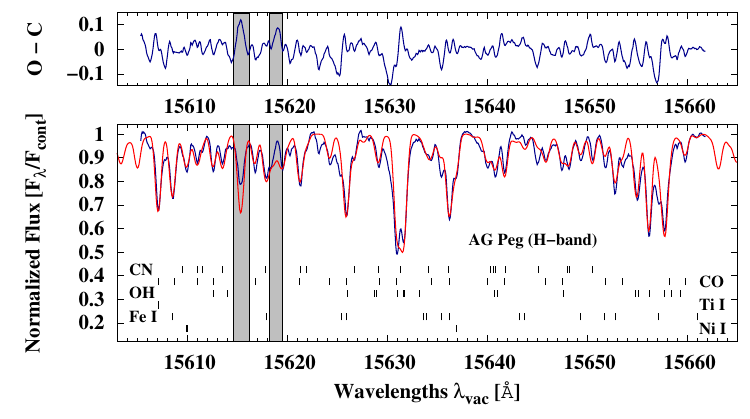}
      \caption{The $H$ band spectrum of AG\,Peg (blue line) and a synthetic
              spectrum (red line) calculated using the final abundances
              (Table\,\ref{T-finAbu}). The grey-shaded areas
              were excluded from calculations by a suitable
              mask.}
         \label{FC27}
   \end{figure}
%

%
   \begin{figure}
   \centering
   \includegraphics[width=\hsize]{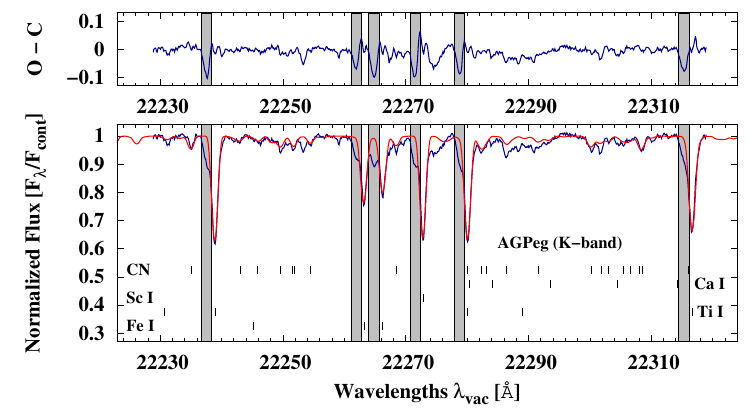}
      \caption{The $K$ band spectrum of AG\,Peg (blue line) and a synthetic
              spectrum (red line) calculated using the final abundances
              (Table\,\ref{T-finAbu}). The grey-shaded areas
              were excluded from calculations by a suitable
              mask.}
         \label{FC28}
   \end{figure}
%

%
   \begin{figure}
   \centering
   \includegraphics[width=\hsize]{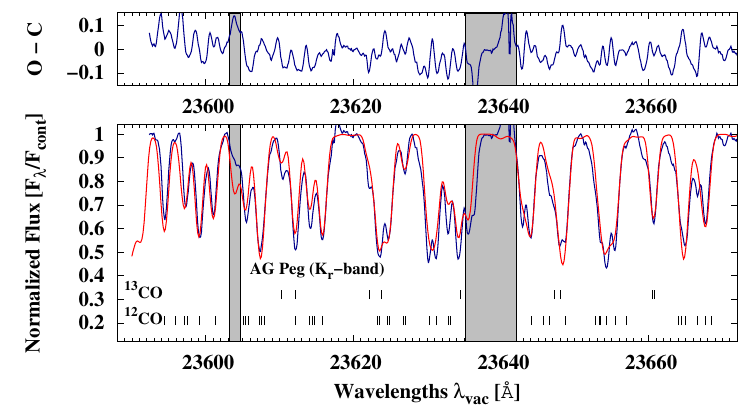}
      \caption{The $K_{\rm r}$ band spectrum of AG\,Peg (blue line) and a
              synthetic spectrum (red line) calculated using the final abundances
              (Table\,\ref{T-finAbu}). The grey-shaded areas
              were excluded from calculations by a suitable
              mask.}
         \label{FC29}
   \end{figure}
%

%
   \begin{figure}
   \centering
   \includegraphics[width=\hsize]{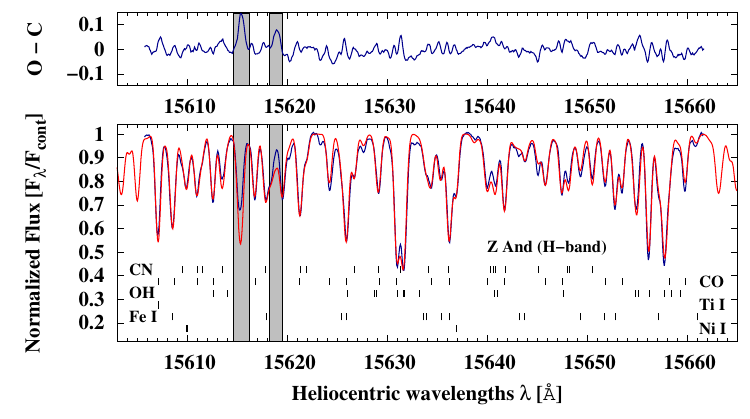}
      \caption{The $H$ band spectrum of Z\,And (blue line) and a synthetic
              spectrum (red line) calculated using the final abundances
              (Table\,\ref{T-finAbu}). The grey-shaded areas
              were excluded from calculations by a suitable
              mask.}
         \label{FC30}
   \end{figure}
%

%
   \begin{figure}
   \centering
   \includegraphics[width=\hsize]{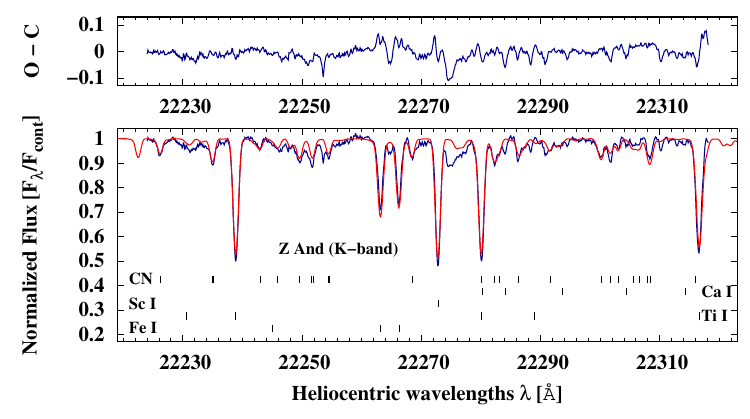}
      \caption{The $K$ band spectrum of Z\,And (blue line) and a synthetic
              spectrum (red line) calculated using the final abundances
              (Table\,\ref{T-finAbu}).}
         \label{FC31}
   \end{figure}
%

%
   \begin{figure}
   \centering
   \includegraphics[width=\hsize]{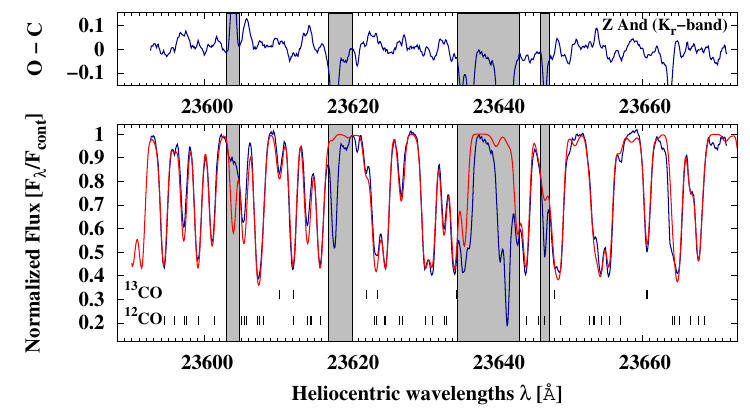}
      \caption{The $K_{\rm r}$ band spectrum of Z\,And (blue line) and a
              synthetic spectrum (red line) calculated using the final abundances
              (Table\,\ref{T-finAbu}). The grey-shaded areas
              were excluded from calculations by a suitable
              mask.}
         \label{FC32}
   \end{figure}
%


\bsp	
\label{lastpage}
\end{document}